\documentclass[article,accept,moreauthors]{Definitions/mdpi} 

\pubvolume{1}
\issuenum{1}
\articlenumber{0}
\pubyear{2023}
\copyrightyear{2023}
\datereceived{ } 
\daterevised{ } 
\dateaccepted{ } 
\datepublished{ } 

\usepackage{cancel}
\usepackage[normalem]{ulem}
\usepackage{xcolor}
\usepackage{soul}

\renewcommand{\papercitation}{}
\renewcommand{\journalname}{}
\fancypagestyle{plain}{\fancyfoot[R]{}\fancyfoot[L]{}}

\Title{Precursors to Topological Phase Transition in Topological Ladders}

\TitleCitation{Precursors to Topological Phase Transition in Topological Ladders}

\Author{Seungju Han and Mahn-Soo Choi *}

\AuthorNames{Seungju Han and Mahn-Soo Choi}

\AuthorCitation{Han, S.; Choi, M.-S.}

\address{%
  Department of Physics, Korea University, Seoul 02841, South Korea}

\corres{Correspondence: choims@korea.ac.kr}

\abstract{We study the coupling of two topological subsystems in distinct topological states, and show that it leads to a precursor behavior of the topological phase transition in the overall system. This behavior is solely determined by the symmetry classes of the subsystem Hamiltonians and coupling terms, and is marked by the persistent existence of subgap states within the bulk energy gap.
By investigating the critical current of Josephson junctions involving topological superconductors, we also illustrate that such subgap states play crucial roles in physical properties of nanoscale devices or materials.}

\keyword{topological insulator;
topological supercondcutors;
zero-energy edge states;
suggap states}

\newcommand\up{\uparrow}
\newcommand\down{\downarrow}
\newcommand\hatc{\hat{c}}
\newcommand\hatH{\hat{H}}
\newcommand\hatV{\hat{V}}

\newcommand{\hlred}[1]{#1}
\newenvironment{revised}{}{}

\begin{document}

\section{Introduction}
In the more traditional frameworks of the first-order or second-order phase transitions, the states of matter have been classified by symmetry \cite{Goldenfeld92a}. For topological materials, the states are determined by topology, more specifically, the topological properties of the momentum space that are independent of deformation \cite{Hasan10a}.
By now, there are a plethora of studies on topological states of matter and physical phenomena resulting from the particular topological states \cite{Qi11a,Chiu16a,Armitage18a,Aguado17}.
The signature property of gapped topological materials, i.e., topological insulators, is the so-called zero-energy states, i.e, the states pinned at the zero energy in the middle of the insulating energy gap and localized at the boundary of the material \cite{Chiu16a}.
Ultimately related to the bulk-boundary correspondence of the topological states, the zero-energy states are almost always presumed to be the only subgap states (if any) in the energy spectra of topological insulators (including topological superconductors). That is, it is generally accepted that a topological insulator is characterized by continua of bulk energies and, in the topologically non-trivial phase, the discrete zero-energy states.

In this work, we \hlred{reveal} a generic situation where a topological insulator retains subgap states of topological origin other than the zero-energy states.
In particular, we show that the interplay between topological states of matter in two coupled \hlred{one-dimensional} topological materials, \hlred{which we call ``topological ladder'' for its schematic structure}, leads to \hlred{distinct subgap states in the topologically trivial phase. The existence of such subgap states proves to be} a genuine \emph{precursor behavior} of the topological phase transition in the overall system. 
\hlred{This} precursor behavior is governed completely by the symmetry classes of the subsystem Hamiltonians and the coupling term, and are characterized by the persistent presence of subgap states inside the bulk gap.
\hlred{In many cases, e.g., the critical current of Josephson junctions involving topological superconductors \cite{haim15,haim15b,Tiira17a,Onder18a,Ren21a,Levajac2023}, 
it is crucial to carefully take into account such subgap states.}
The effects of these additional subgap states get more pronounced in nanostructures or low-dimensional materials such as nanowires \cite{liu12,huang18,Bagrets12a,Aguado17,Penaranda18a}.

\begin{revised}
We note that alongside the zero-energy states, the subgap Andreev bound states \cite{huang18c} and near-zero-energy states \cite{liu12,huang18c,Bagrets12a,prada12,Kells12a,degottardi13,ojanen13,roy13,rainis13,adagideli14,stanescu14,jelena15,sanjose16,mtdeng16,liu17,ptok17,fleckenstein18,avila19,moore18,moore18c,tudor19,woods19,chen19,awoga19,Prada2020,ahn21,marra22,fukaya23,Driel23,haim15b,haim15,hessDloss23,Huaiqiang18a,reegDloss2017,anselmetti19}
that emerge in the topologically \emph{trivial} phase have attracted considerable attention as well. This is because such Andreev bound states display features akin to the true zero-energy states within a topologically non-trivial phase, i.e., the Majorana zero modes \cite{Alicea11a,Bagrets12a,rainis13,adagideli14,moore18} that play an essential role for topological quantum computation\cite{Kitaev06b,Alicea11a}, and may lead to substantial experimental controversies \cite{Zhang21a}. 
Numerous scenarios for manipulating Andreev bound states have been investigated, encompassing factors such as nonuniform effective chemical potential \cite{Kells12a,roy13,ojanen13,degottardi13,stanescu14,jelena15,moore18c}, coupling to a quantum dot \cite{liu17,ptok17,prada12}, multi-band interactions \cite{woods19,awoga19,ahn21,fukaya23}, and partially separated Andreev bound states \cite{moore18c,tudor19,Prada2020,marra22}.
Interestingly, the {intrinsic finite-energy Andreev bound states} found in a numerical study based on both continuum and tight-binding models \cite{huang18c} provide a direct evidence of our mechanism of subgap states of topological origin.
\end{revised}

\section{Model}
\label{sec::model}

We consider a coupled system of two bulk-gapped topological subsystems governed by Hamiltonian of the form
\begin{align}
\label{topham1}
\hat{H}_{\text{total}} = \hat{H}_{A} + \hat{H}_{B} + \lambda\hat{V} ,
\end{align}
where $\hat{H}_{A}$ and $\hat{H}_{B}$ represent the Hamiltonians of the individual subsystems $A$ and $B$ and $\hat{V}$ describes the coupling between the two subsystems.
Each subsystem is assumed to be in a topologically non-trivial phase.
Here, we have introduced an explicit parameter $\lambda$ to indicate the coupling strength.
We will investigate the regime where $\lambda$ is turned on from zero to a small but finite value.
\hlred{
To be specific, we primarily focus on two parallel (or even spatially overlapping) one-dimensional topological subsystems, denoted as $\hat{H}_{A}$ and $\hat{H}_{B}$. The coupling $\hat{V}$ is between the corresponding parallel sites belonging to different systems. To stress this schematic coupling structure, we call the overall system a \emph{topological ladder}.}

Later, we will show that the coupling $\hatV$ leads to a generic interplay between topologically distinct states in the two topological materials $A$ and $B$, which gives rise to persistent subgap states inside the bulk gap.
As a brief overview, let us consider a specific case where two subsystems belonging to the BDI symmetry class.
When the subsystems are decoupled ($\lambda = 0$), the system belongs to the
BDI $\oplus$ BDI symmetry class. Suppose that both subsystems are in the topological phase. Then, there are edge states associated with each of the subsystems localized at the boundaries.
We find that with weak coupling $\lambda$, there
must be genuine subgap states (including the zero-energy states) inside the bulk gap in the total system.
This is regardless of whether the total system is in the topological or
trivial phase. However, it is instructive to distinguish the
two cases. If the total system is in topologically trivial state after turning
on $\lambda$, then the subgap states must have finite energy (still inside the
bulk gap). In this case, the persistence of the subgap states (well separated
from the bulk spectrum) is guaranteed energetically because the weak coupling
cannot lift the degenerate states from the subsystems into the bulk continuum
with a finite gap.  On the other hand, if the total system is in the
topological phase, then the subgap states are at the zero energy. Obviously,
in this case, they are protected for the topological reasons.
In both cases, unless $\lambda$ is too large compared with the bulk-energy gap, the subgap states
are localized at the boundaries of the system.

\begin{figure}
\begin{centering}
\begin{tabular}{c}
\includegraphics[width=6.6cm]{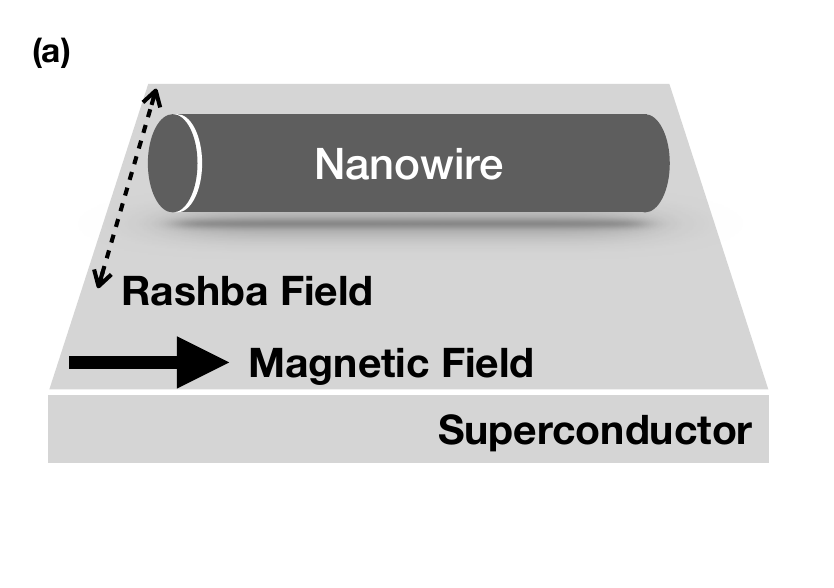}\quad
\includegraphics[width=6.6cm]{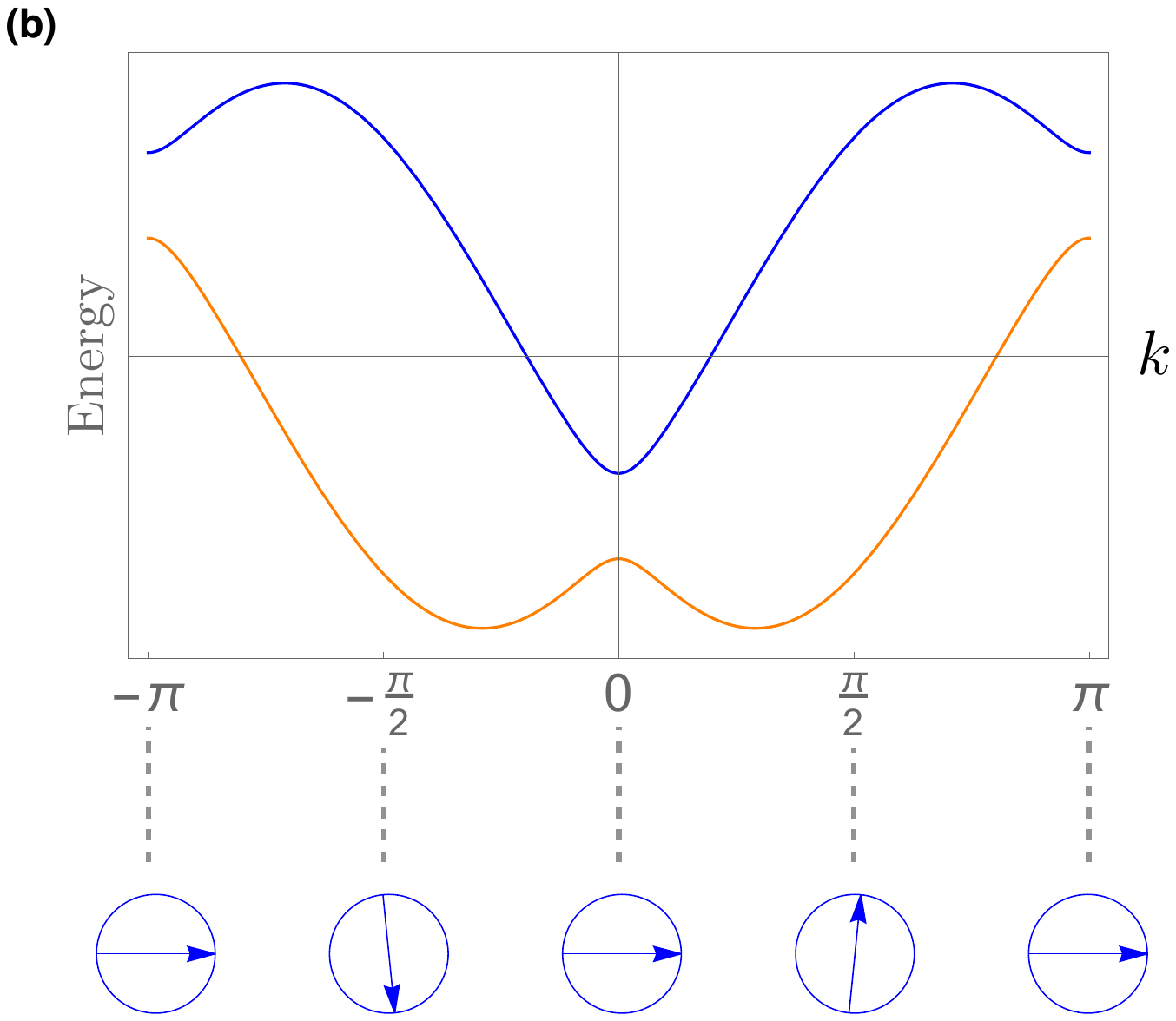}
\par\end{tabular}
\end{centering}
\caption{(color online)
  (a)
  A schematic of a nanowire system with an external magnetic field on a superconductor.
  The dotted double-headed arrow indicates the direction of the magnetic field induced by the Rashba-type spin-orbit coupling on the nanowire.
  (b)
 A dispersion relation without superconductivity, $E^{\pm}_{so}(k)$.
Blue arrows indicate the directions of the effective magnetic field at a given momentum.
\hlred{They correspond to the spin direction of the electron in the blue band.}
  At $k=0,\pm\pi$, the effective field is aligned to the direction of the Zeeman field (positive $x$-axis) due to the absence of the Rashba field.
  At $k=\pm \pi /2$, the effective field is most affected by the Rashba field, which has an opposite direction for opposite momentum.}
\label{fig::intro}
\end{figure}

\subsection{Nanowire as an independent two-band model}
\label{sec::model:nanowire}

In many physical systems, the interplay mentioned above is not
immediately apparent. The most prominent example is the nanowire with
spin-orbit coupling in proximity to a superconductor and in an external magnetic
field as shown in Fig.~\ref{fig::intro} (a).
This system is widely recognized as a physical realization of the theoretical model of Kitaev chain, and is known to demonstrate the topological phase transition and associated Majorana edge states predicted in the Kitaev chain \cite{Kitaev01a,Oreg10a,Lutchyn10a}.

Throughout this work, we will argue that in fact, the nanowire system involves two topological subsystems (more precisely, two bands in topologically non-trivial phase as elaborated below) and the interplay between them gives rise to a precursor behavior \hlred{in terms of subgap states} before the topological phase transition.
\hlred{Such subgap states were also noted in a previous numerical study \cite{huang18c}, but were recognized merely as intrinsic Andreev bound states with no reference to topological states.}

Here, we briefly review a common picture where the two \emph{uncoupled} bands in the nanowire system leads to the topological phase transition predicted in the Kitaev chain. Later, we will point out that the residual coupling between the two subsystems plays an important role.
\begin{revised}
The tight-binding Hamiltonian for the system is given by
\begin{multline}
\label{hamiltonianposition}
\hat{H}
= - t \sum_{x=1}^{N} \sum_{s=\uparrow,\downarrow} \hat{c}_{x+1,s}^{\dagger} \hat{c}_{x,s} 
+ \frac{\Delta_{so}}{2} \sum_{x=1}^{N-1} \left(  \hat{c}_{x+1,\uparrow}^{\dagger} \hat{c}_{x,\downarrow} + \hat{c}_{x+1,\downarrow}^{\dagger} \hat{c}_{x,\uparrow} \right)
\\ {}
+ \Delta_Z \sum_{x=1}^{N} \hat{c}_{x,\uparrow}^{\dagger} \hat{c}_{x,\downarrow}
+ \Delta_{sc} \sum_{x=1}^{N} \hat{c}_{x,\uparrow}^{\dagger} \hat{c}_{x,\downarrow}^{\dagger}
+ h.c. - \mu \sum_{x=1}^{N} \sum_{s=\uparrow,\downarrow} \left( \hat{c}_{x,s}^{\dagger} \hat{c}_{x,s} - \frac{1}{2} \right).
\end{multline}
The momentum-space representation in the periodic boundary condition leads to
\end{revised}
\begin{align}
\label{hamiltonian}
\hat{H}_\mathrm{bulk}
&= \frac{1}{2}\sum_{k} \hat{\Psi}^{\dagger}_{k}
H_\mathrm{bulk}(k)
\hat{\Psi}_k	\end{align}
with
\begin{equation}
H_\mathrm{bulk}(k) =
\begin{bmatrix}
H_{so}(k) & \Delta_{sc}\sigma_0 \\
\Delta_{sc}^*\sigma_0 & -H_{so}^*(-k)
\end{bmatrix} ,
\end{equation}
where
\begin{equation}
\label{hso}
H_{so}(k) =
\left\{ -\mu\hlred{ -  2t  \cos k} \right\} \sigma_0+ \Delta_Z \sigma_x  + \Delta_{so} \sin k \sigma_y
\end{equation}
is the single-particle Rashba Hamiltonian and
\begin{equation}
\hat{\Psi}^{\dagger}_{k} =
\begin{bmatrix}
\hat{c}^{\dagger}_{k\uparrow} &
 \hat{c}^{\dagger}_{k \downarrow} &
\hat{c}_{-k\uparrow} &
\hat{c}_{-k\downarrow} 
\end{bmatrix}
\end{equation}
is the spinor field in the Nambu notation, $\mu$ is the chemical potential of the system, $\Delta_{so}$ is the Rashba type spin-orbit coupling along $y$-direction, $\Delta_Z$ is the Zeeman field along $x$-direction, and $\Delta_{sc}$ is the proximity-induced superconducting pairing potential.
Without loss of generality, all the parameters are assumed to be positive (with proper choice of gauge if necessary).
Band width \hlred{$4t$} and chemical potential $\mu$ are typically the largest energy scales in this system, and accordingly, we will assume that both are much larger than both the spin-orbit coupling \hlred{$\Delta_{so}$} and superconducting pairing potential $\Delta_{sc}$ ($t,\mu > \Delta_{so},\Delta_{sc}$).
\hlred{In any discussion of nanowire throughout the paper, we take the \emph{effective} Fermi energy $\mu+2t$ as the unit scale of energy.}

As well known \cite{Oreg10a}, the above system exhibits a topological phase
transition at the critical Zeeman field
\begin{math}
\Delta_Z^* := \sqrt{\hlred{\left(\mu+2t\right)^2} + \Delta_{sc}^2} .
\end{math}
For Zeeman field $\Delta_Z$ greater
than $\Delta_Z^*$, the system is in a topological regime hosting a pair
of Majoranas with one Majorana at each end.
We call this the `main' topological phase transition to distinguish it from
similar transitions in individual subsystems.

The typical picture describing the main topological phase transition starts with identifying two bands.
The single-particle Rashba Hamiltonian in Eq.~\eqref{hso} gives rise to the two momentum-locked spin bands due to the spin-orbit coupling, i.e.,
\begin{equation}
\hatH_{so}(k) = \hlred{\sum_\pm \hatH_{so}^\pm(k)} = \sum_\pm E_{so}^\pm(k)\,\hatc_{k,\pm}^\dag\hatc_{k,\pm}.
\end{equation}
where the dispersions $E_{so}^\pm$ are given by 
\begin{equation}
E^{\pm}_{so}(k)
= -\mu \hlred{-2t \cos k} \pm \sqrt{\Delta_Z^2 + \Delta_{so}^2 \sin^2 k }
\end{equation}
The two electron operators $\hatc_{k,\pm}$ on the two bands $\pm$ are related to the bare electron operators $\hatc_{k,\up/\down}$ via the following unitary transformation
\begin{equation}
\label{paper::eq:3}
\begin{bmatrix}
\hat{c}_{k,+}  \\ \hat{c}_{k,-}
\end{bmatrix}
=  U_k
\begin{bmatrix}
\hat{c}_{k,\uparrow}  \\ \hat{c}_{k,\downarrow}
\end{bmatrix} \,,\quad
U_{k} := \frac{1}{\sqrt{2}}
\begin{bmatrix}
e^{i\phi_k} &1 \\
-e^{i\phi_k}  & 1
\end{bmatrix},
\end{equation}
where $\phi_k$ is the angle between the directions of the Zeeman field and the effective magnetic field that \hlred{an} electron with momentum $k$ feels, i.e.,
\begin{math}
\sin\phi_k = \Delta_{so} \sin k / \Delta_Z ;
\end{math}
see the blue arrows in Fig.~\ref{fig::intro} (b).
Their typical dispersion relations are as shown in Fig.~\ref{fig::intro} (b). In these bands, the spin quantization axes are oriented parallel and anti-parallel, respectively, to the effective magnetic field consisting of the external Zeeman field $\Delta_Z$ in the $x$-direction and the momentum-dependent spin-orbit field $\Delta_{so}\sin{k}$ in the $y$-direction.

\begin{figure*}
\begin{centering}
\begin{tabular}{cc}
\includegraphics[width=60mm]{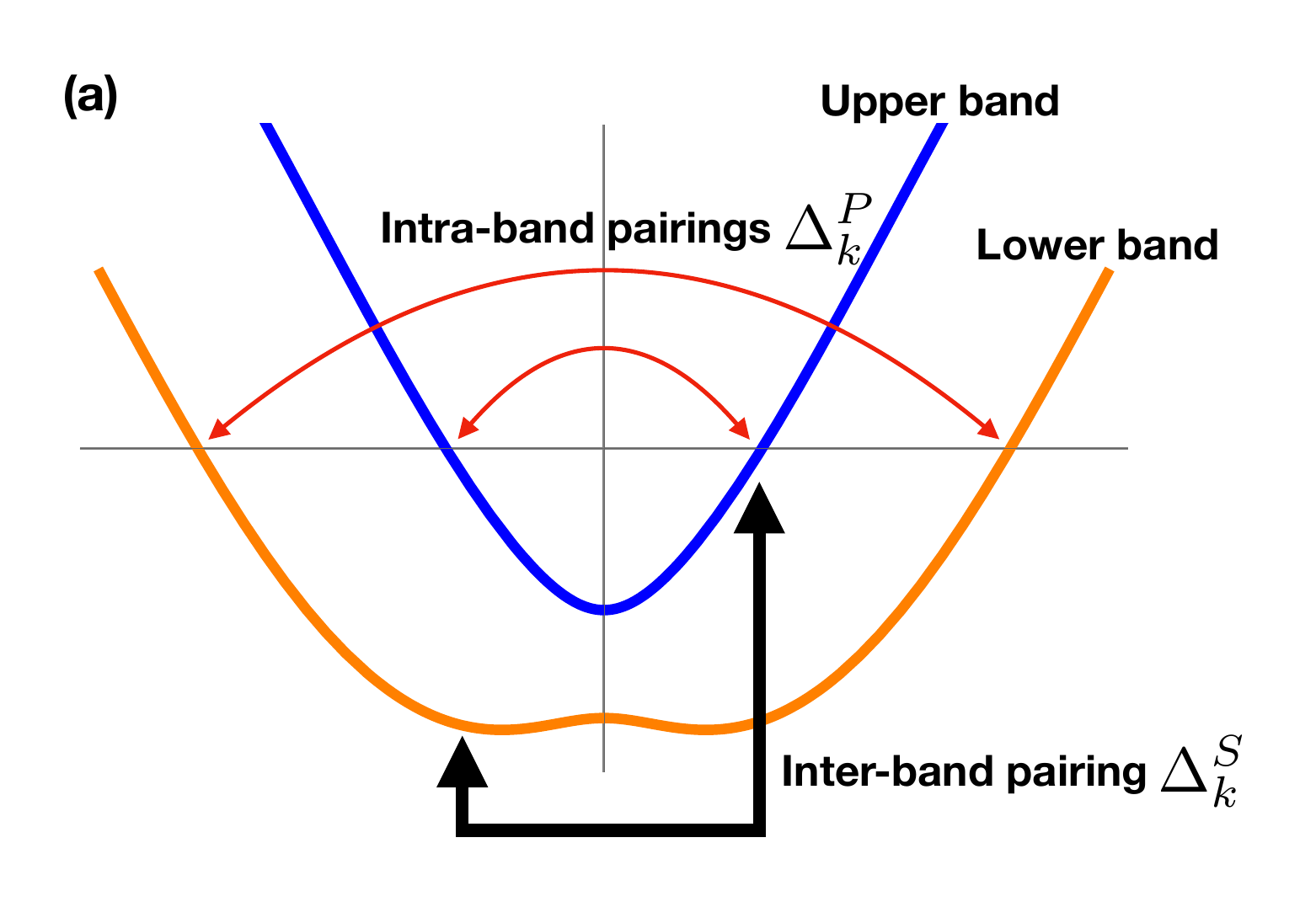} &
\includegraphics[width=60mm]{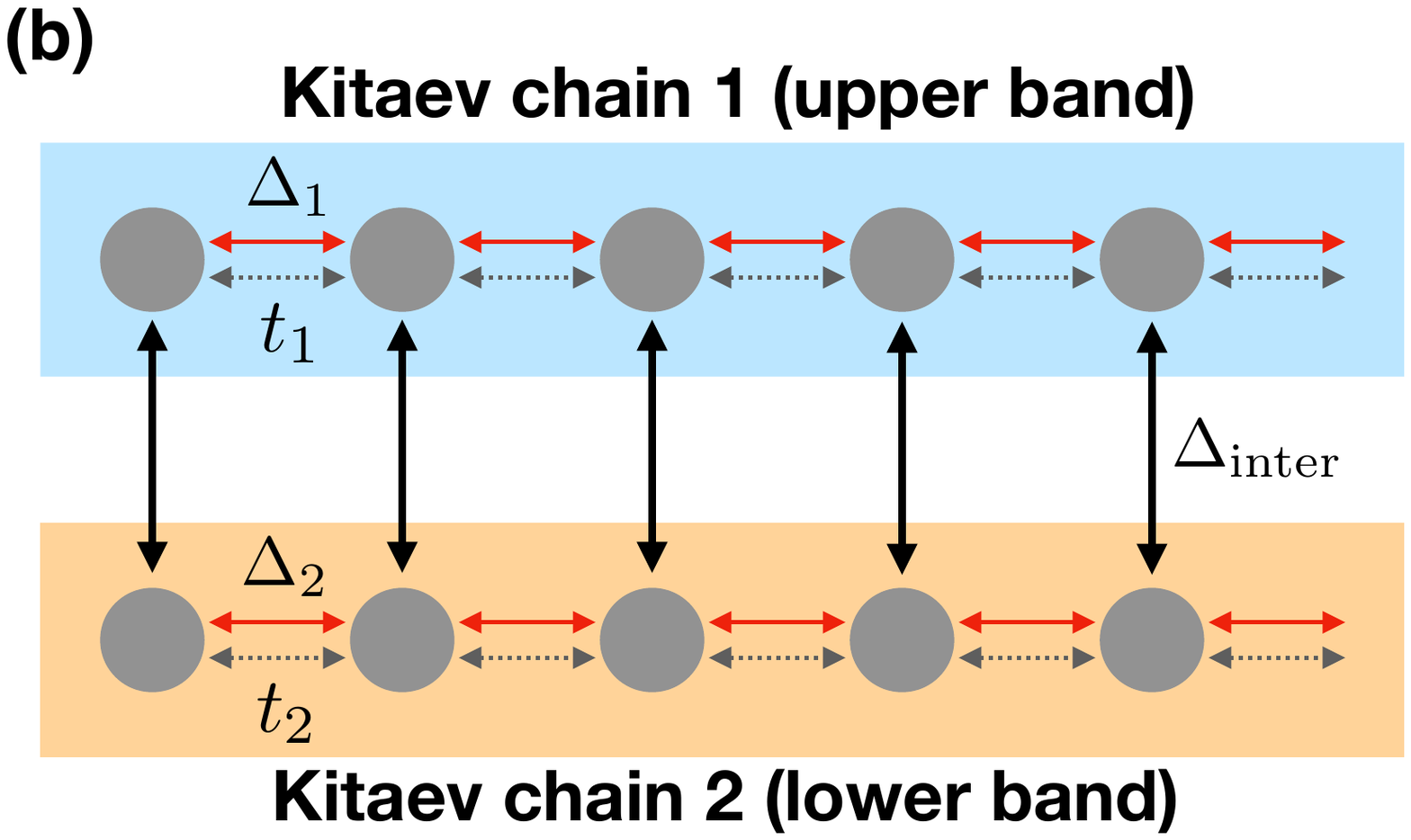}
\end{tabular}
\par\end{centering}
\caption{
(color online)
Relation between the nanowire \hlred{(a)} and coupled Kitaev chains \hlred{(b)}.
\hlred{Panel (a) describes the dispersion relation without superconductivity of the nanowire, with superconducting pairings represented by arrows.}
Curved red arrows \hlred{illustrate} pairings within the same band\hlred{, while} the black arrow indicates pairing between different bands.
\hlred{Panel (b) describes the tight-binding model of the coupled Kitaev chains.}
\hlred{The blue (orange) stripe containing gray disks represents a Kitaev chain that corresponds to the upper (lower) band of the nanowire.}
Black vertical arrows depict the parametric coupling between two Kitaev chains, which corresponds to the inter-band pairing of the nanowire.
\hlred{In the (approximate) mapping between the nanowire and coupled Kitaev chains, the inter-band pairing $\Delta_{k}^{S}$ (intra-band pairing $\pm \Delta_{k}^{P}$, respectively) corresponds to the inter-chain pairing $\Delta_{\text{int}}$ (intra-chain pairing $\Delta_\alpha for \alpha=1,2$, respectively).
}
}
\label{fig::analogy:pairing}
\end{figure*}

\begin{figure*}
\begin{centering}
\begin{tabular}{cc}
\includegraphics[width=60mm]{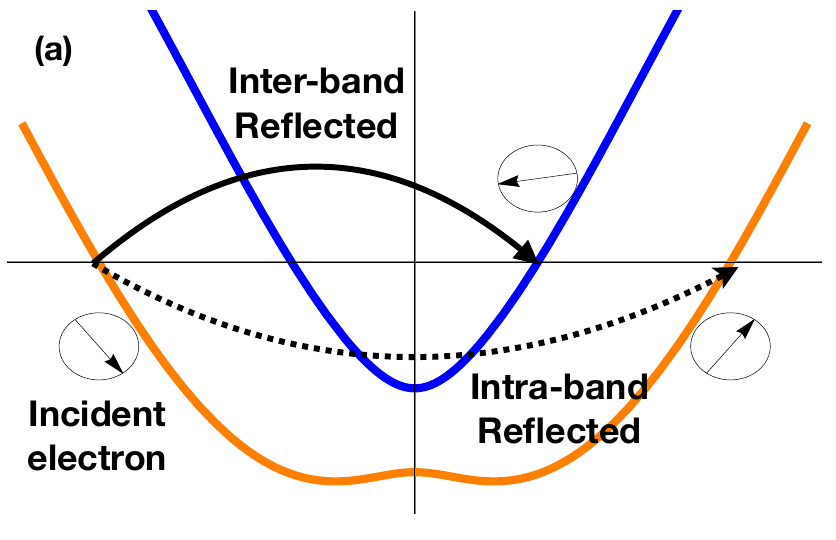} &
\includegraphics[width=60mm]{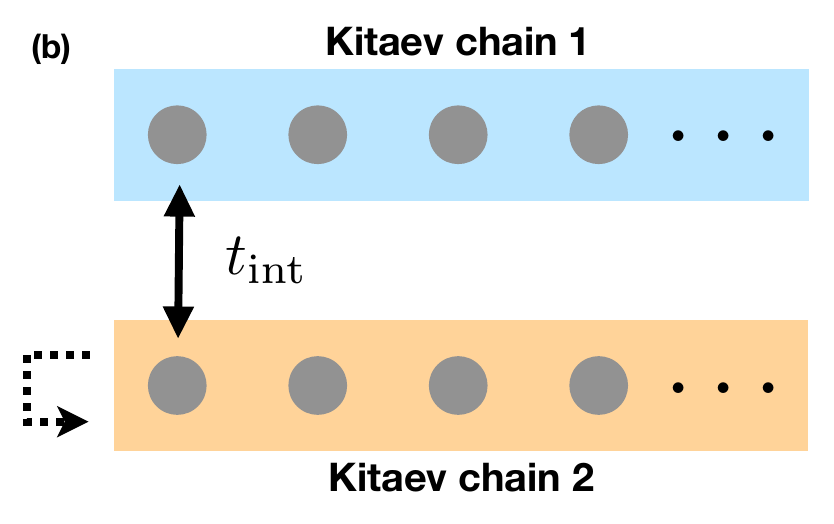}
\end{tabular}
\par\end{centering}
\caption{
(color online)
(a) \hlred{A} schematic illustration of reflections at the Fermi energy. The arrows within black circles represent spin directions, and the blue and orange curves display the energy dispersions for both bands.
The dotted-curve arrow \hlred{indicates} intra-band reflection, and the solid-curve arrow represents inter-band reflection.
(b) Inter-band \hlred{hopping (with amplitude $t_\mathrm{int}$)} and intra-band reflection in coupled Kitaev chains \hlred{are depicted by solid and dotted arrows, respectively}.}
\label{fig::analogy:reflections}
\end{figure*}

\hlred{Due to the locking of spin to momentum, each of the two bands $E_{so}^\pm$ described above can be treated mathematically as representing a \emph{spinless} fermion.}
Therefore, turning on the proximity-induced superconducting pairing potential $\Delta_{sc}$ \hlred{effectively leads} to a $p$-wave pairing within each band.
Overall, the bulk Hamiltonian for the nanowire in Eq.~\eqref{hamiltonian} can be \hlred{approximately} described by
\begin{equation}
\label{paper::eq:4}
\hatH_\mathrm{bulk} \approx \hatH_+ + \hatH_-
\end{equation}
with
\begin{equation}
\label{paper::eq:1}
\hat{H}_\pm :=\frac{1}{2} 
\sum_k\hat\Psi_\pm^\dagger(k)
\begin{bmatrix}
E_{so}^\pm(k) & \pm i\Delta_k^{P} \\
\mp i\left(\Delta_k^{P}\right)^* & -E_{so}^\pm(k)
\end{bmatrix}\hat\Psi_\pm(k) \,,
\end{equation}
where
\begin{math}
\Delta_k^{P} := \Delta_{sc}\sin\phi_k
\end{math}
is the strength of the $p$-wave pairing potentials within
individual bands $\pm$, and
we have introduced new spinor fields
\begin{equation}
\label{eq:basis}
\hat\Psi_\pm(k) :=
\begin{bmatrix}
\hat{c}^{\dagger}_{k,\pm} &
\hat{c}_{-k,\pm}
\end{bmatrix}^T .
\end{equation}
In this simplified picture, \hlred{the two bands may individually be in either the topological or trivial phase,}
depending on the values of the chemical potential $\mu$ and the external Zeeman splitting $\Delta_Z$.
For example, if $\Delta_Z<\Delta_Z^*$ (with \hlred{$\mu<2t$}), then both bands cross the Fermi level and would be in the topologically non-trivial phase, leading to two Majorana edge states (from respective bands) at each end of the nanowire.
Since two Majoranas occupying the same location merge into one Dirac fermion, the topological state of the nanowire in this case is trivial. On the other hand, if
\begin{math}
\Delta_Z^*<\Delta_Z<\sqrt{(\hlred{\mu-2t})^2 + \Delta_{sc}^2},
\end{math}
then the upper band $E_{so}^+$ is off the Fermi level while the lower band $E_{so}^-$ remains crossing the Fermi level, leaving one Majorana edge state at each end, and hence the nanowire is in the topologically non-trivial phase.
This explains the main topological phase transition at $\Delta_Z=\Delta_Z^*$.

In reality, however, the two bands are coupled with each other and cannot be treated separately as in the above argument.

\subsection{Nanowire as a coupled two-band model}

The coupling between the two momentum-locked spin bands \hlred{$\hatH_{so}^{\pm}$} arises
for two different reasons. One is the bulk effect.
Note that due to
the external Zeeman field, the spin quantization axes for opposite momenta
are not exactly opposite.
Considering that superconducting pairing occurs between electrons with  opposite spins and momenta, this
leads to off-diagonal components
of the pairing in the
momentum-locked spin basis.
More specifically, the paring potential can be divided into
two parts, the intra-band pairing, $\Delta^P$, already seen in Eq.~\eqref{paper::eq:1} and inter-band pairing,
\begin{math}
\Delta_k^S := \Delta_{sc}\cos\phi_k,
\end{math}
depending on whether it is diagonal or off-diagonal; see Fig.~\ref{fig::analogy:pairing} (a).
This leads to a more complete description of the nanowire Hamiltonian in Eq.~\eqref{hamiltonian} in the form 
\begin{equation}
\hat{H}_\mathrm{bulk} = \hatH_+ + \hatH_- + \hat{V} ,
\end{equation}
where $\hatH_\pm$ describe the two bands $\pm$ as defined in Eq.~\eqref{paper::eq:1} and
\begin{equation}
\label{paper::eq:2}
\hat{V} := \frac{1}{2} \sum_{\pm} \hat\Psi_{\pm}^\dag(k)
\begin{bmatrix}
0 & \pm\Delta_k^S \\
\mp \left(\Delta_k^S\right)^*& 0
\end{bmatrix}\hat\Psi_{\mp}(k)
\end{equation}
is responsible for the inter-band coupling.

The other source of the coupling between two bands is the reflection at the boundary of the nanowire.
This coupling originates in the spin mismatch between left-going and right-going electrons in the same band.
When a left-going electron in the lower band, for example, reflects at the boundary, it cannot be reflected exclusively into the lower band due to the spin conservation (see Fig.~\ref{fig::analogy:reflections} (a)).
There should be a reflected wave in the upper band as well. 
This can be clearly seen when we express
the annihilation operators, $\hat{c}_{x,\uparrow}$ and $\hat{c}_{x,\downarrow}$,
for electrons with spin up and down, respectively,
in terms of the operators $\hat{c}_{y,\pm}$ for electrons with the momentum-locked spins $\pm$,
\begin{equation}
\label{positionrep}
\begin{bmatrix} \hat{c}_{x,\uparrow}	\\ \hat{c}_{x,\downarrow} \end{bmatrix}
= \sum_y F(x-y)
\begin{bmatrix}
\hat{c}_{y,+}	\\
\hat{c}_{y,-}
\end{bmatrix} ,
\end{equation}
where
\begin{equation}
F(x-y) := \frac{1}{N}\sum_k e^{ik(x-y)} U_k^{\dagger}
\end{equation}
with unitary matrix $U_k$ defined in Eq.~\eqref{paper::eq:3}.
Since the spin rotation matrix $U_k$ depends on $k$, $F(x-y)$ is not
diagonal in general.
This means that even a simple boundary condition in position space, such as an open or hard boundary, induces reflections from one band to the other.

\subsection{Coupled Kitaev Chains}
\label{sec::twocoupledkitaevchains}

Before we discuss the effects of the inter-band coupling on the topological properties of the nanowire system,
to avoid the unnecessary complexity and to focus on the main physics,
we examine the same effects in a toy model that has the key features of nanowire.
Our toy model is composed of two coupled Kitaev chains as schematically depicted in Fig.~\ref{fig::analogy:pairing} (b) and Fig.~\ref{fig::analogy:reflections} (b), its Hamiltonian reads as
\begin{equation}
\label{twokitaev}
\hat{H}_{\text{CKC}} =
\hat{H}_{\text{KC,1}} + \hat{H}_{\text{KC,2}}	
+ \hat{V}_{\text{D}} +
\hat{V}_{\text{R}}.
\end{equation}
Here, the isolated upper ($\alpha=1$) and lower ($\alpha=2$) Kitaev chains are described by
\begin{align}
\label{eq:noreflection}
\hat{H}_{\text{KC},\alpha}
&= -\hlred{t_\alpha}\sum_{x=1}^{N-1}
\hat{c}^{\dagger}_{x+1,\alpha} \hat{c}_{x,\alpha}
+ \hlred{\Delta_\alpha}\sum_{x=1}^{N-1}\hat{c}^{\dagger}_{x+1,\alpha}
\hat{c}^{\dagger}_{x,\alpha} + h.c.
\nonumber \\
&- \mu_\alpha\sum_{x=1}^{N}
\left( \hat{c}^{\dagger}_{x,\alpha} \hat{c}_{x,\alpha} - \frac{1}{2} \right) ,
\end{align}
where
\begin{math}
\hat{c}^{\dagger}_{x,\alpha}
\end{math}
\begin{math}
(\hat{c}_{x,\alpha})
\end{math}
is a fermion creation (annihilation) operator at site $x$,
$t_\alpha > 0$ the hopping amplitude,
$\Delta_\alpha$ the pairing potential,
and $\mu_\alpha$ the chemical potential on chain $\alpha$.
We set $\hlred{2}t_\alpha > |\mu_\alpha| $, such that without any coupling between the chains, both chains are in the topological phase.
The term
\begin{equation}
\hat{V}_{\text{D}}
= \Delta_{\text{int}} \sum_{x=1}^N\hat{c}^{\dagger}_{x,1}
\hat{c}^{\dagger}_{x,2} + h.c.
\end{equation}
is responsible for the coupling due to the inter-chain pairing potential, allowing the Kitaev chains to be geometrically viewed as a topological ladder [see Figure~\ref{fig::analogy:pairing} (b)].
Whereas the term
\begin{equation}
\hat{V}_{\text{R}}
= \left( -t_{\text{int}}\hat{c}^{\dagger}_{1,1} \hat{c}_{1,2}
+ t_{\text{int}}\hat{c}^{\dagger}_{N,1} \hat{c}_{N,2}\right)+ h.c.
\end{equation}
describes the reflection from one chain to the other at the boundaries ($x=1,N$) with reflection amplitude $\mp t_{\text{int}}$ [see  Fig.~(\ref{fig::analogy:reflections}) (b)].

We take the toy model in Eq.~\eqref{twokitaev}
as a simplified model of realistic nanowires,
and the connection between the two models is already clear from corresponding and resembling terms.
However, some of the simplifications is worth mentioning.
In the nanowire model, the inter-band pairing,
\begin{math}
\Delta_k^S = \Delta_{sc}\cos\phi_k
\end{math}
in Eq.~\eqref{paper::eq:2}, 
depends on momentum $k$. However, the complex phase (including signs) does not change as the momentum $k$ varies because $-\pi/2 < \phi_k < \pi/2$.
\hlred{Therefore, we consider the inter-chain pairing potential $\Delta_{\text{int}}$ to be constant in $k$, as depicted in }Fig.~\ref{fig::analogy:pairing} (b).
The relative phases of all three pairing potentials, $\Delta_1$, $\Delta_2$ and \hlred{$\Delta_{\text{int}}$} cannot be eliminated by a U(1) gauge transformation, but the relative phases of any two of them can be fixed by choosing a proper gauge. Throughout the work, we fix the complex phases of intra-chain pairing $\Delta_1$ and $\Delta_2$ to $\pi$ and $0$, respectively, i.e., $\Delta_1 < 0$ and $\Delta_2 > 0$.

\section{Persistent subgap states}

\subsection{Coupled Kitaev Chains} 
\label{sec::twokitaev}

As the first examples of physical implications of the interplay of two subsystems in topologically distinct states, we \hlred{initially} investigate the toy model of coupled Kitaev chains described in Eq.~(\ref{twokitaev}).
As we assumed in Section~\ref{sec::twocoupledkitaevchains}, both Kitaev chains are in the topological phase when decoupled, \hlred{with} each chain hosting a Majorana bound state at its each end.
It will be instructive to regard the inter-chain pairing $\Delta_{\text{int}}$ and hopping $t_{\text{int}}$ as independent parameters.
When we eventually compare this toy model to the nanowire model, the inter-chain pairing $\Delta_{\text{int}}$ must be real-valued \hlred{and the complex phase of inter-chain hopping $t_\text{int}$ should be determined based on the nanowire model.}
\hlred{In this subsection,} we lift \hlred{these constraints}, allowing \hlred{them} to take an \emph{arbitrary} complex phase, to investigate the broader impacts of coupling on the symmetry classes of the total system.

\begin{figure*}
\includegraphics[width=0.31\textwidth]{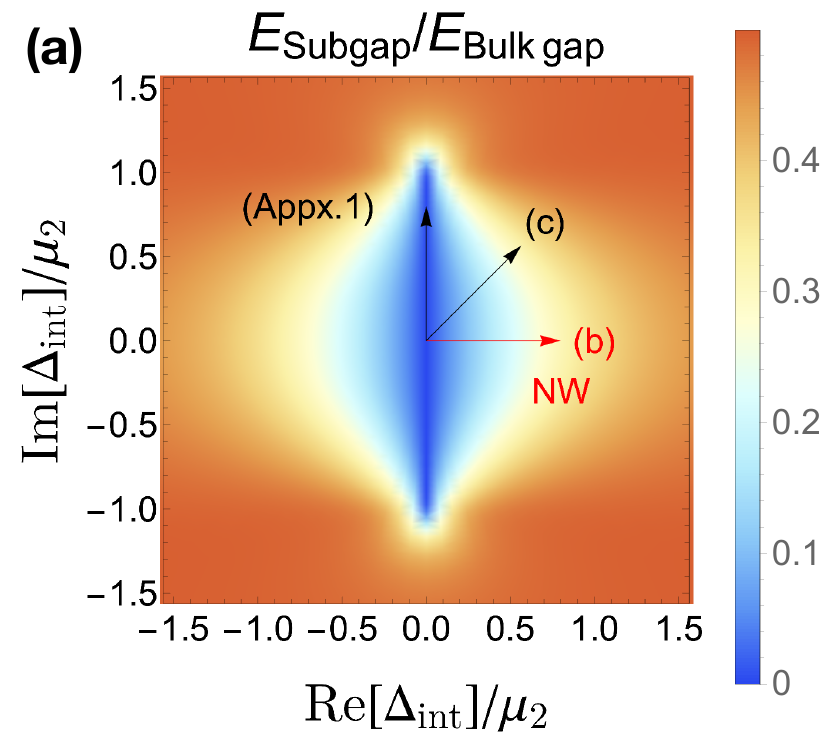} 
\includegraphics[width=0.31\textwidth]{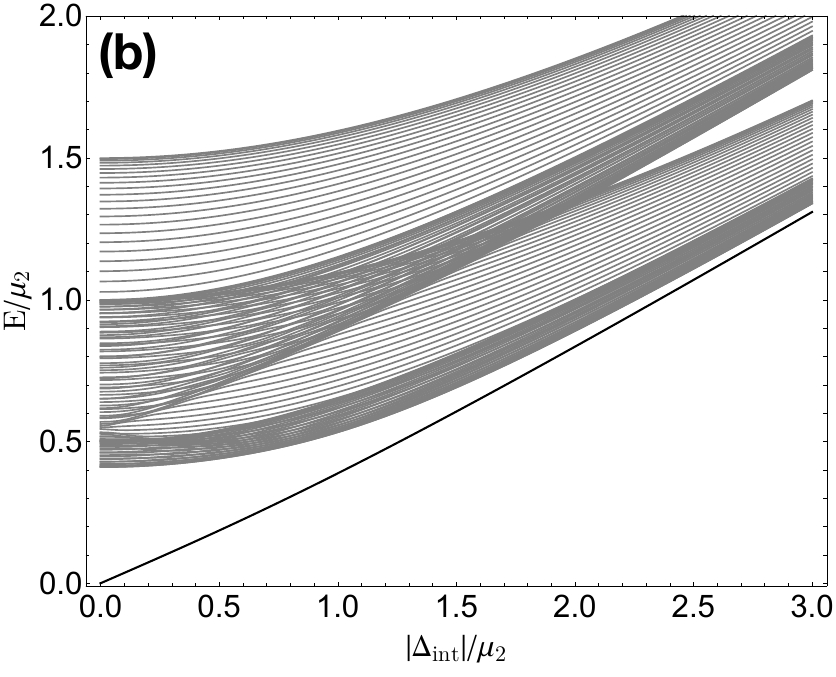}
\includegraphics[width=0.31\textwidth]{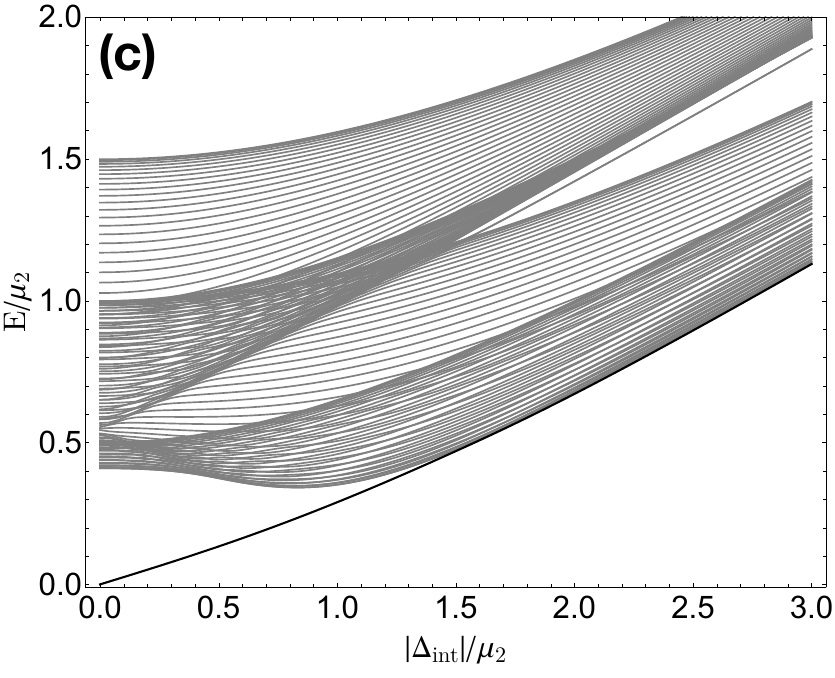} 
\caption{
(color online)
Energy profiles of the two coupled Kitaev chains model \hlred{are depicted} as a function of inter-chain parametric pairing, $\Delta_{\text{int}}$.
Inter-chain reflection, i.e., inter-chain hopping at boundaries, is turned off.
\hlred{Only the positive energy levels are shown as the system has particle-hole symmetry.}
(a) Ratio between the subgap-state energy and bulk gap energy.
The subgap-state energy is measured from the Fermi level, \hlred{so} that the ratio $E_{\text{Subgap}}/E_{\text{Bulk gap}}$ converges to $0.5$ when the subgap energy is merged into bulk.
\hlred{The arrow marked with (Appx.1) is to indicate that the implied direction is discussed in} the appendix.
(b,c) Energy levels along the arrows in panel (a), i.e., complex phase of the pairing is fixed to \hlred{$0$ and $\pi/4$} in (b) and (c), respectively.
The pure real $\Delta_{\text{int}}$ corresponds to the nanowire model.
Parameters: The number of sites for each chain $N=60$,
on-site energy $\mu_1=0$, $\mu_2=1$ (unit energy scale),
intra-chain pairings $\Delta_{1}=-\Delta_{2}=-\hlred{\mu_2/2}$,
hoppings \hlred{$t_1=t_2=\mu_2$.}
}
\label{fig::twokitaev:pairing}
\end{figure*}

First, let us turn on only the inter-chain pairing;
\hlred{$\Delta_\mathrm{int}$} $\neq0$ and $t_{\text{int}} = 0$.
The ratio between bulk gap energy and subgap energy is shown in Fig.~\ref{fig::twokitaev:pairing} (a)\hlred{, obtained by diagonalizing Eq.~}\eqref{eq:noreflection}.
When the coupling is zero, $\Delta_{\text{int}} = 0$, the two Majoranas at each end exist at zero energy(by ignoring the small coupling proportional to $e^{-L}$), such that the ratio between the bulk gap and subgap energy $E_{\text{Subgap}} / E_{\text{Bulk gap}}$ is zero.
As we turn on the coupling, the subgap energy starts to change.
Note that if the inter-chain pairing can have an arbitrary complex phase, 
\hlred{the coupled system belongs to symmetry class D with its associated} topological invariant $\mathcal{Z}_2$.
It means that at least one of the two Majorana bound states at one end should disappear.
However, as it is well known, a single Majorana bound state cannot disappear through \hlred{a small} interaction with bulk states because of the finite bulk gap energy.
Therefore, the system should be in a topologically trivial phase, such that the two Majorana bound states are to interact with each other and be merged into a Dirac state with its energy within the bulk gap, which is shown in Fig.~\ref{fig::twokitaev:pairing} (a).
The nanowire analogy result is shown in Fig.~\ref{fig::twokitaev:pairing} (b).
As the parameter, $\Delta_{\text{int}}$, changes along the positive real axis, we have nonzero subgap state energy within the bulk gap\hlred{, which is a precursor of the \emph{main} topological phase transition discussed in Sec.~}\ref{sec::model:nanowire}

Note that an accidental degeneracy could arise, i.e., an accidental symmetry may prevent the two Majoranas from interacting with each other.
Under the exclusion of the inter-band hopping, it happens when the inter-chain pairing is pure imaginary, $\Delta_{\text{int}} = \pm i |\Delta_{\text{int}} |$.
In this case, the subgap states are fixed at zero energy until the coupling becomes sufficiently large to close the band gap, and the subgap states disappear through the band touching.
The detailed discussion of this regime will be presented in Appendix Sec. \ref{sec::appendix1} with relevant figures in Fig.~\ref{fig::twokitaev:appendix}.

\begin{figure*}
\includegraphics[width=0.31\textwidth]{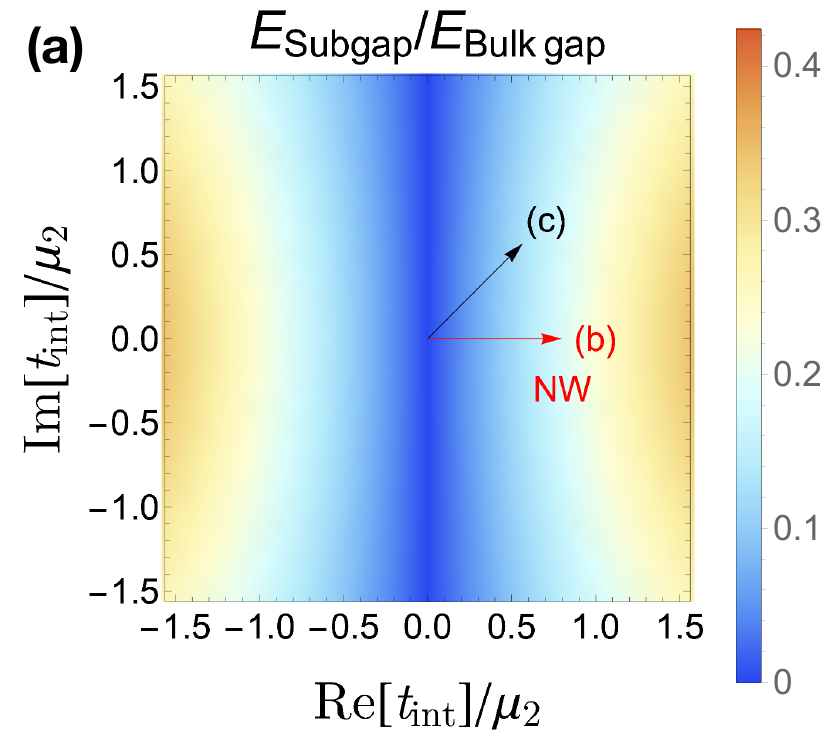} 
\includegraphics[width=0.31\textwidth]{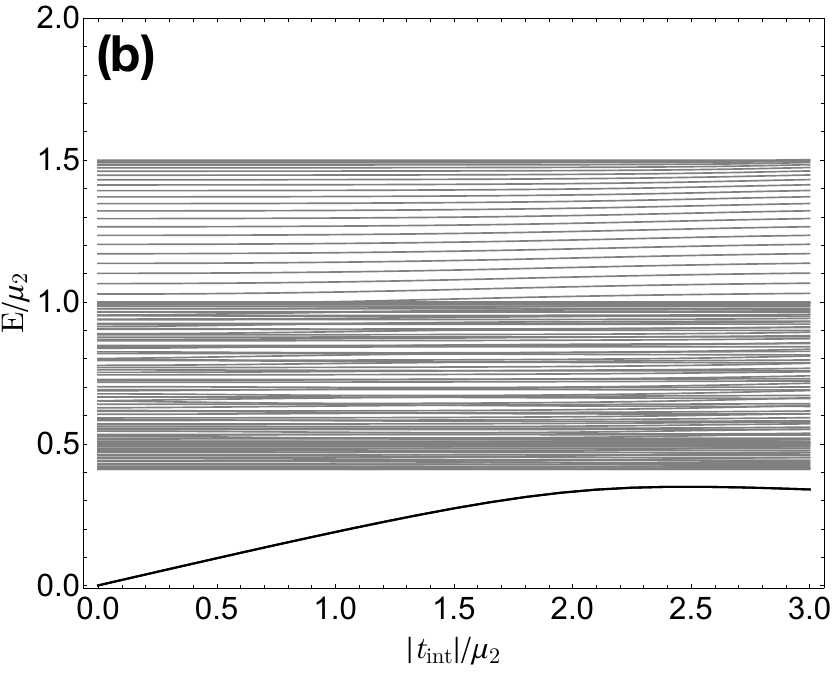}
\includegraphics[width=0.31\textwidth]{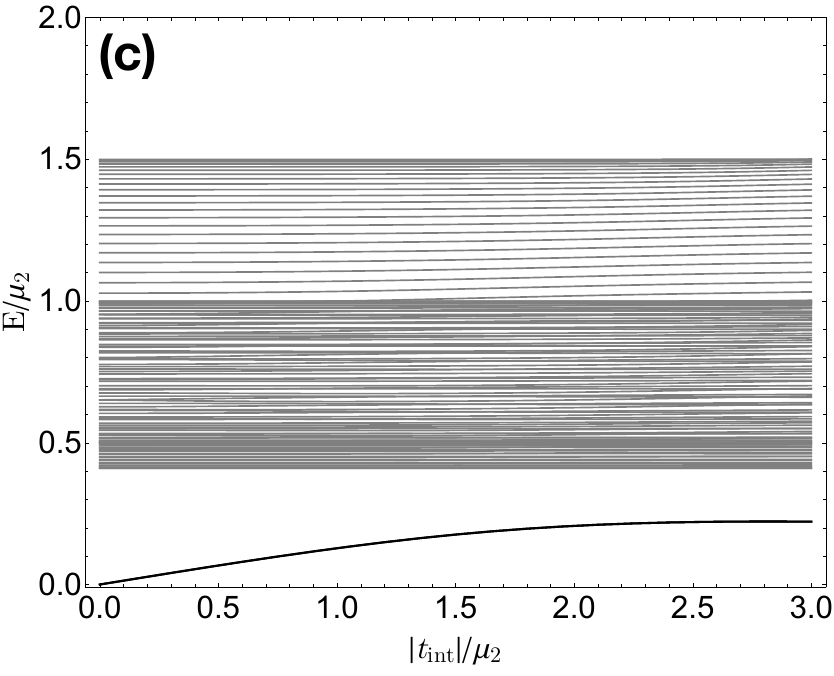} 
\caption{
(color online)
Energy profiles of the two coupled Kitaev chains model as a function of inter-chain hopping, $t_{\text{int}}$.
The inter-chain pairing is set to zero, $\Delta_{\text{int}}=0$.
\hlred{Only the positive energy levels are shown.}
(a) Ratio between the subgap state energy and bulk gap energy.
The subgap energy is measured from the Fermi level.
(b,c) Energy levels along the arrows in panel (a) \hlred{are depicted.} \hlred{The} complex phase of the hopping is fixed to \hlred{$0$ and $\pi/4$} in (b) and (c), respectively.
The pure real $t_{\text{int}}$ corresponds to the nanowire model.
Parameters: The number of sites for each chain $N=60$,
on-site energy $\mu_1=0$, $\mu_2=1$ (unit energy scale),
intra-chain pairings $\Delta_{1}=-\Delta_{2}=-\hlred{\mu_2/2}$,
hoppings \hlred{$t_1=t_2=\mu_2$}.
}
\label{fig::twokitaev:ref}
\end{figure*}

Second, we set \hlred{the inter-chain pairing} to zero and consider the inter-chain reflection as the only coupling between the chains.
\hlred{It is worth noting that geometrically, the system does not conform to the structure of topological ladders, i.e., it lacks rungs in the middle.
Nonetheless, we are intrigued by the influence of the inter-chain reflection on Majorana fermions, as both are localized at the endpoints.}
The results are presented in Fig.~\ref{fig::twokitaev:ref}.
Unlike the varying $\Delta_{\text{int}}$ case, the inter-chain hopping exists only at the boundaries, such that the bulk band does not show any significant changes.
Besides this difference, the overall behavior of the energy of the subgap state can be seen within the similar context.
The set of Hamiltonians with the arbitrary complex number $t_{\text{int}}$ is in the symmetry class D, so the system is in a topologically trivial phase for the same reason as mentioned above.
Starting with zero energy states at $t_{\text{int}}=0$, we have finite subgap energy well separated from the bulk band.

\begin{figure*}
\includegraphics[width=0.31\textwidth]{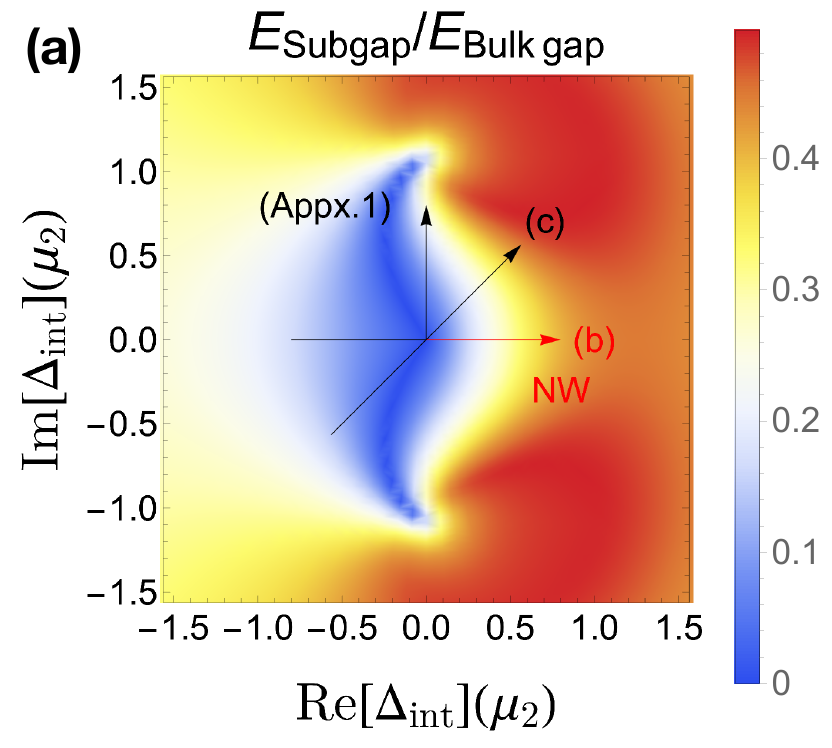} 
\includegraphics[width=0.31\textwidth]{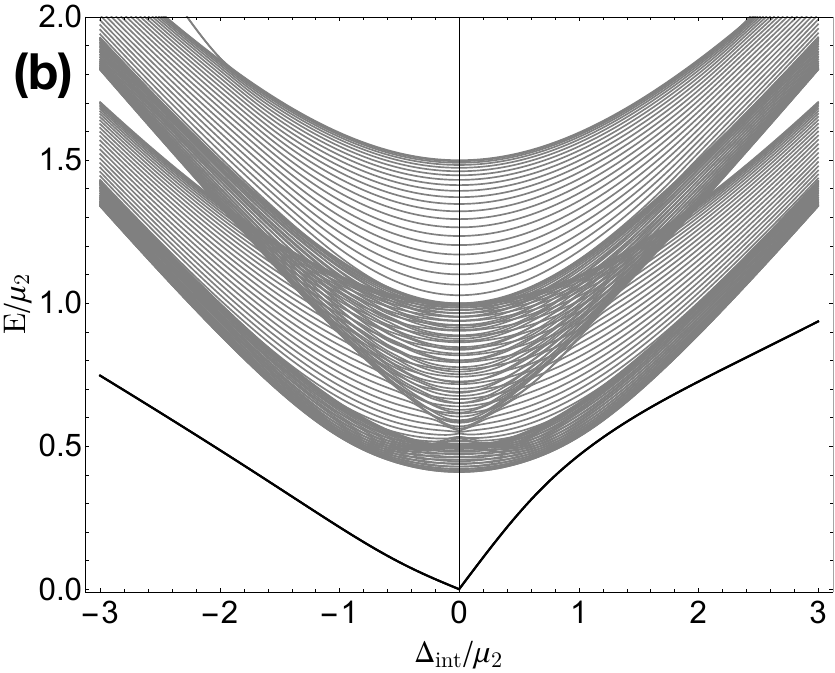} 
\includegraphics[width=0.31\textwidth]{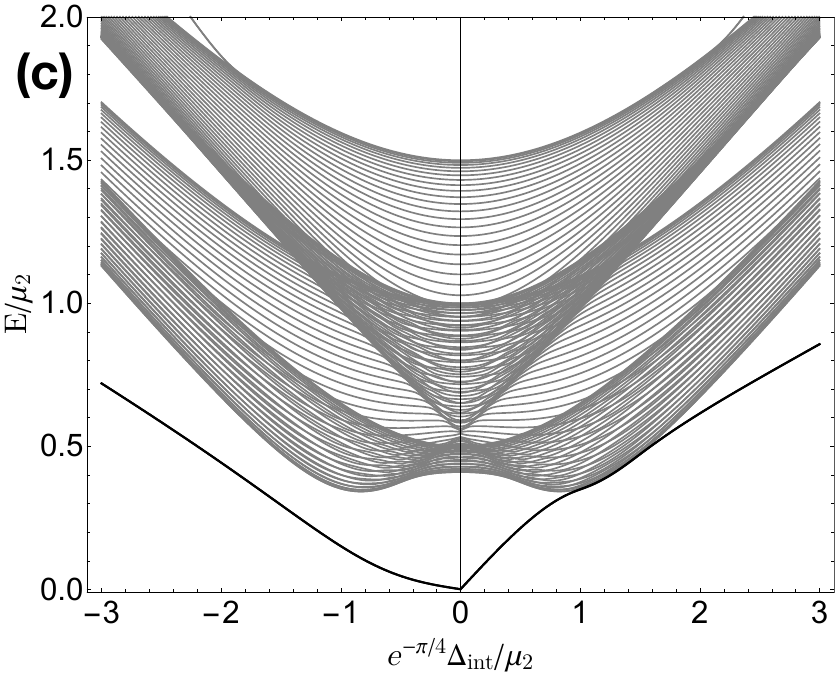}
\caption{
(color online)
Energy profiles of the two coupled Kitaev chains model as a function of inter-chain pairing, $\Delta_{\text{int}}$, with inter-chain \hlred{reflection} $t_{\text{int}} = |\Delta_{\text{int}}|$ at the boundaries.
(a) Ratio between the subgap state energy and bulk gap energy.
\hlred{The inset arrow, marked with (Appx.1), is} for the appendix.
(b,c) Energy levels as function of $\Delta_{\text{int}}$ with its complex phase are indicated as an arrow in panel (a).
Parameters: The number of sites for each chain $N=60$,
on-site energy $\mu_1=0$, $\mu_2=1$ (unit energy scale),
intra-chain pairings $\Delta_{1}=-\Delta_{2}=-\hlred{\mu_2/2}$,
hoppings \hlred{$t_1=t_2=\mu_2$.}
}
\label{fig::twokitaev:total}
\end{figure*}

Finally, in Fig.~\ref{fig::twokitaev:total}, we \hlred{consider} both inter-chain pairing and \hlred{reflection} simultaneously.
We set the pure real inter-chain \hlred{reflection} $t_{\text{int}} =|\Delta_{\text{int}}|$ and vary the pairing with an arbitrary complex phase as we did previously.
Unlike the previous results, the interplay between two couplings, $t_{\text{int}}$ and $\Delta_{\text{int}}$, makes \hlred{the} subgap energy now depend on the complex phase of $\Delta_{\text{int}}$ more asymmetrically.

\subsection{Nanowire}

\begin{figure*}
\begin{centering}
\includegraphics[width=8cm]{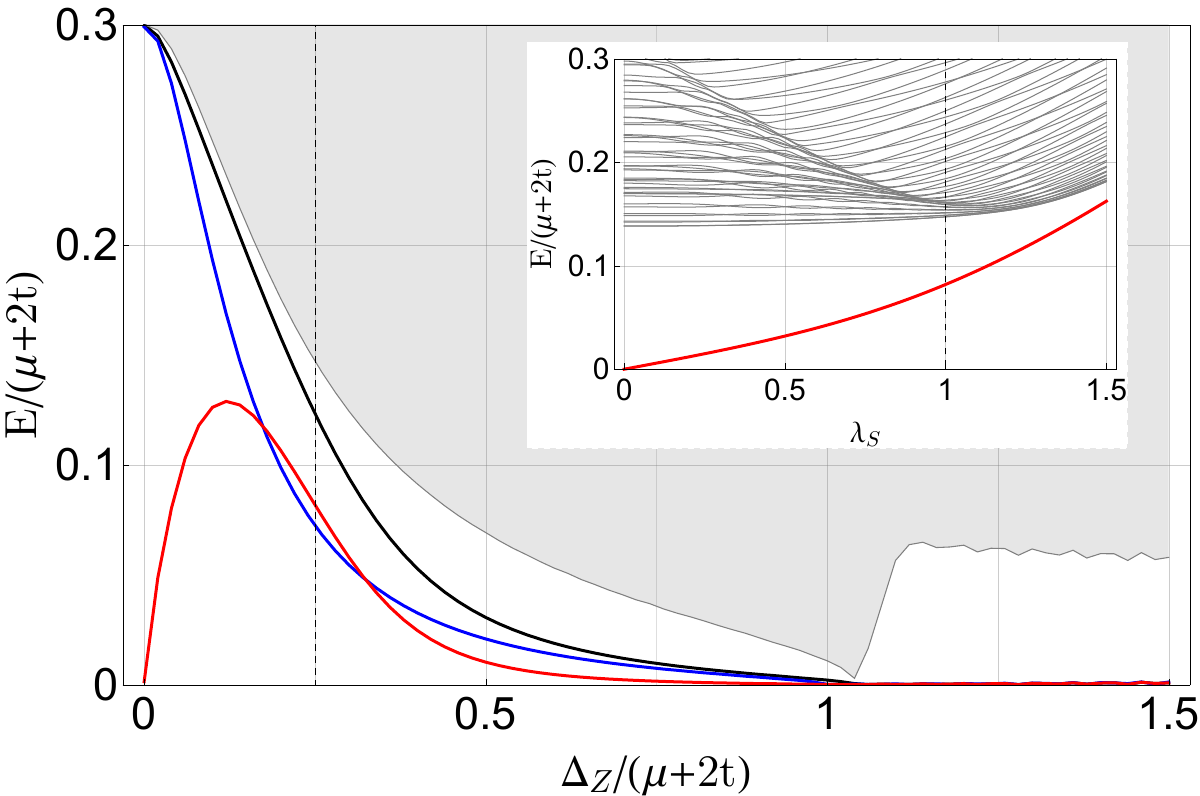}
\par\end{centering}
\caption{
(color online)
The subgap energy \hlred{under} different conditions and \hlred{the} bulk energy levels of the nanowire as a function of the Zeeman field is depicted.
\hlred{The} gray area indicate\hlred{s} bulk state energy, and black line represents the subgap state energy.
The red line \hlred{illustrates} the subgap energy when inter-band reflection is forbidden.
The blue line \hlred{shows} the subgap energy without the inter-band pairing, \hlred{that is,} inter-band reflection is the only coupling between bands.
The vertical dashed line is \hlred{a reference point} for the inset figure.
\hlred{We used tight-binding model with $N=500$ sites, $t=7/4$, $\mu = -5/2$ (setting the unit energy scale $\mu+2t=1$),
$\Delta_{sc} = 0.3$, and $\Delta_{so} = 0.3$.}
The critical Zeeman field is approximately $1.044$.
(Inset) Energy level versus modified inter-band pairing strength $\Delta^S_k \rightarrow \lambda^S \Delta^S_k$ for all $k$\hlred{, with } inter-band reflection turned off.
The vertical dashed line indicate\hlred{s} $\lambda_S=1$.
\hlred{Parameters for the inset are consistent with those in the main plot at $\Delta_Z=0.25$.}
}
\label{fig::nanowire}
\end{figure*}

In this section, we \hlred{turn our attention to a more realistic model: the nanowire.
The black solid line and the gray area in Fig.~\ref{fig::nanowire}, obtained by finding eigenvalues of Eq.~\eqref{hamiltonian}, depict the subgap energy and the bulk band of the nanowire system, respectively, as functions of the Zeeman field.}
As mentioned earlier in Sec.~\ref{sec::model:nanowire},
when the Zeeman field is larger than the critical field $\Delta_Z^*$, the system is in the \hlred{topologically non-trivial} phase.
Majorana bound states exist at the zero energy \hlred{when we disregard} the exponentially small coupling $e^{-N}$ \hlred{with respect to the system size $N$, physically, the distance between the Majorana states at opposite edges}.
With the Zeeman field smaller than the critical field, we find the subgap states (black curve) originates in the interplay of two Majoranas from upper and lower bands, respectively.
\hlred{We set the number of site as $N=500$, which is regarded sufficient to avoid the finite-size effects.}

Before we \hlred{delve into} the result,
\hlred{let's separately consider inter-band pairing and reflection, just as we did in the previous section.}

First, \hlred{we examine} the case with truncated inter-band reflection.
To eliminate the inter-band reflection, we express the Hamiltonian in momentum space using the momentum-locked spin basis depicted in Eq.~\eqref{eq:basis}.
\hlred{Subsequently, we perform the inverse Fourier transform, writing the position-space operator $\hat{c}_{x_i,\pm}$ as shown in} Eq.~\eqref{positionrep}.
Because of the non-local nature of this basis, the open boundary can no longer be \hlred{expressed} in a simple form.
Instead of using the usual open boundary condition, as \hlred{our focus is on} ignoring the inter-band reflection, we \hlred{directly} impose a new open boundary condition on this basis.
\hlred{This approach yields} the red curve in Fig.~\ref{fig::nanowire}, \hlred{representing} the subgap-state energy without the inter-band reflection.
When the Zeeman field is zero, there \hlred{is} no inter-band pairing due to the exact spin matching between opposite-spin electrons with the opposite momentum\hlred{, resulting in the convergence of the subgap energy to zero.}
As the Zeeman field increases, the inter-band pairing also increases, \hlred{leading to a} finite subgap-state energy.
\hlred{Therefore, it becomes challenging to provide a qualitative description beyond the peak near $\Delta_Z \approx 0.1\mu$.}

To verify \hlred{the proper truncation of inter-band reflection} and show the finite subgap energy originates \hlred{from} inter-band pairing, we calculate energy levels as a function $\lambda_S$, where $\lambda_S$ for a modifier in front of the inter-band pairing, $\Delta^S_k \rightarrow \lambda_S \Delta^S_k$ for all $k$.
See the inset of Fig.~\ref{fig::nanowire}.
As the parameter $\lambda_S$ approaches zero, there is no coupling at all between the two bands, resulting in the subgap state energy approaching zero, consistent with our observations in the two coupled Kitaev chains model.

\hlred{Next, let us consider the scenario where the only coupling between the two bands is through inter-band reflection.}
The result is obtained by \hlred{imposing} the usual open boundary condition to the bulk model with $\lambda_S=0$.
See the blue curve in Fig.~\ref{fig::nanowire}, where the subgap energy solely \hlred{arises} from the inter-band reflection.
When the Zeeman field \hlred{is absent}, the direction of the effective magnetic field is \hlred{solely} determined by the Rashba field. \hlred{Given that} the direction of the Rashba field is opposite for electrons with opposite momentum,
total reflection \hlred{occurs} between the two bands\hlred{, resulting in substantial} inter-band coupling and a large subgap energy that merges into the bulk.

In summary, we \hlred{observe that the nanowire hosts subgap states inthe parameter regime prior to the occurrence of the topological phase transitions, and identify its two origins, inter-band pairing and reflection, by conceptually modeling the nanowire system as a topological ladder.}

\hlred{The possible physical impacts of this precursor behavior prior to topological phase transitions is explored below by focusing on the inter-band pairing through modified pairing potential $\lambda_S\Delta_{\text{int}}$ and keeping the discussions of the inter-band reflection on a qualitative level.}

\section{Critical current}

In this section, we now present the physical implication of the \hlred{aforementioned} subgap states on the critical current of the nanowire junction.

We calculate the zero-bias critical current by using linear response theory for the nanowire connected to \hlred{either} a normal superconductor or a topological superconductor\hlred{, examining its dependence on} the Zeeman field and temperature.
Both the normal and topological superconductors, as well as the nanowire, are \hlred{characterized} by tight-binding models Eq.~\eqref{hamiltonian} with appropriate parameters.
We \hlred{consider} a tunnel junction, expressed by the tunneling Hamiltonian:
\begin{equation}
\label{eq:cctunnel}
H_{\text{tunnel}} = \hlred{t_{\text{tunnel}}} \sum_\sigma \hat{c}^{\dagger}_{L,\sigma} \hat{c}_{R,\sigma} + h.c..
\end{equation}
\hlred{Here,} $\hat{c}^{\dagger}_{L(R),\sigma}$ represents the  electron creation operator for the junction side $L(R)$\hlred{, at one end, and associated} with a spin $\sigma$. The term $t_{\text{tunnel}}$ is a tunneling amplitude.
Without loss of generality, we \hlred{designate} $R$ as the nanowire and $L$ \hlred{as either a normal or topological} superconductor.

To find the contribution of the subgap state to the critical current, we divide the critical current into two \hlred{components: the contribution from localized states and the remaining contribution.
This division is facilitated by expressing Eq.~\eqref{eq:cctunnel} in terms of the eigenstates of  one-dimensional system.}
The meaning of these division will be presented below.

\begin{figure*}
\begin{centering}
\includegraphics[width=8cm]{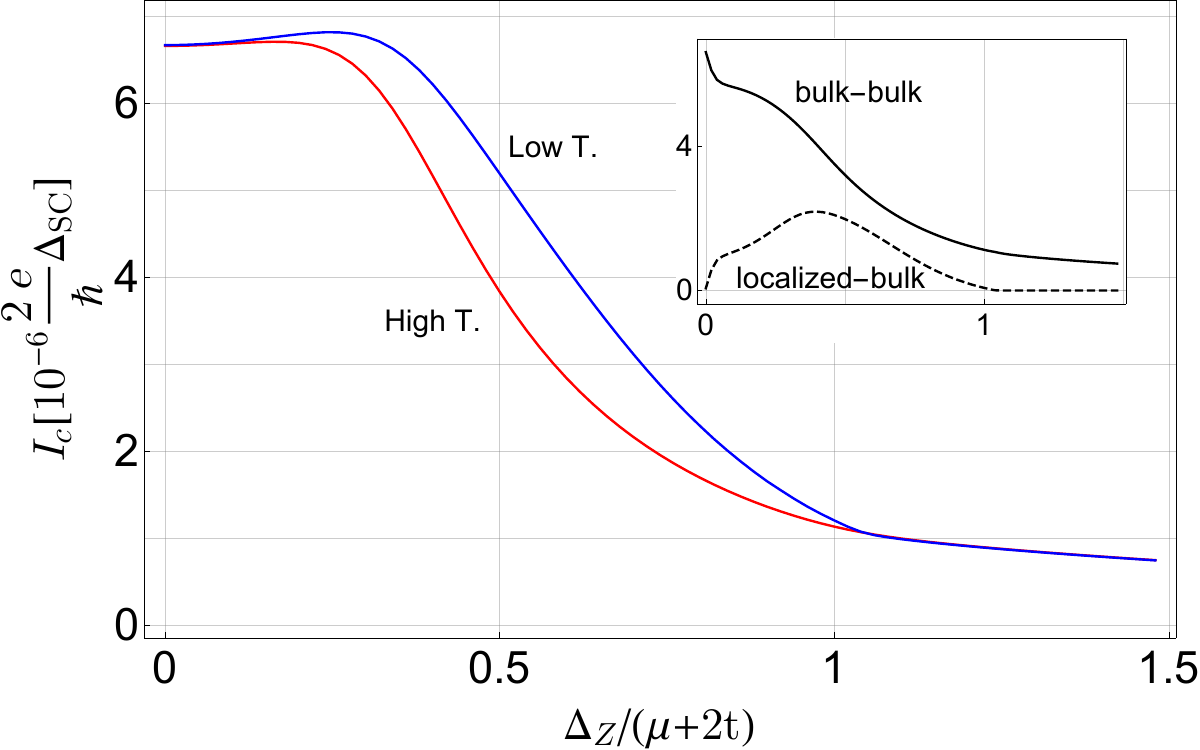}
\par\end{centering}
\caption{
(color online)
Zero bias critical current
$I_c/I_c^{0}$, where $I_c^0:=2e\Delta_{sc}/\hbar$ is the critical current in the short-junction limit,
between a normal superconductor and the nanowire obtained by the tight-binding model is presented.
\hlred{In both the nanowire and the normal superconductor, the parameters are $t=7/4$, $\mu = -5/2$, resulting in the band bottom of $-\mu-2t =  -1$.
The spin orbit coupling $\Delta_{so} = 0.3$ exists only in the nanowire.
Tunneling amplitude is $t_{\text{tunnel}} = 0.02$.
Temperatures are $k_B T = 0.04, 0.005$.}
(Inset)
Critical current through the bulk states of the normal superconductor and the nanowire(Solid line) and through the subgap state at the nanowire and bulk states of the normal superconductor(Dashed line) with the low temperature is given.
Axis units are the same as used in the main plot.
}
\label{fig::cc:nsnw}
\end{figure*}

\hlred{In the case of} the normal superconductor-nanowire junction(NS-NW), the normal superconductor \hlred{lacks} any localized state.
\hlred{Consequently,} as shown in Fig.~\ref{fig::cc:nsnw}, the critical current can be divided into two terms: bulk-bulk and localized-bulk.

\hlred{A discernible contribution to the critical current emerges from the localized state. 
It's noteworthy that, when the Zeeman field surpasses the critical value, indicating that the nanowire is in the topological phase, no supercurrent flows through the localized state (Majorana bound state)} \cite{Zazunov12a}.
The finite contribution of the localized state \hlred{is observable exclusively} in the topologically trivial regime.

\begin{figure*}
\begin{centering}
\includegraphics[width=8cm]{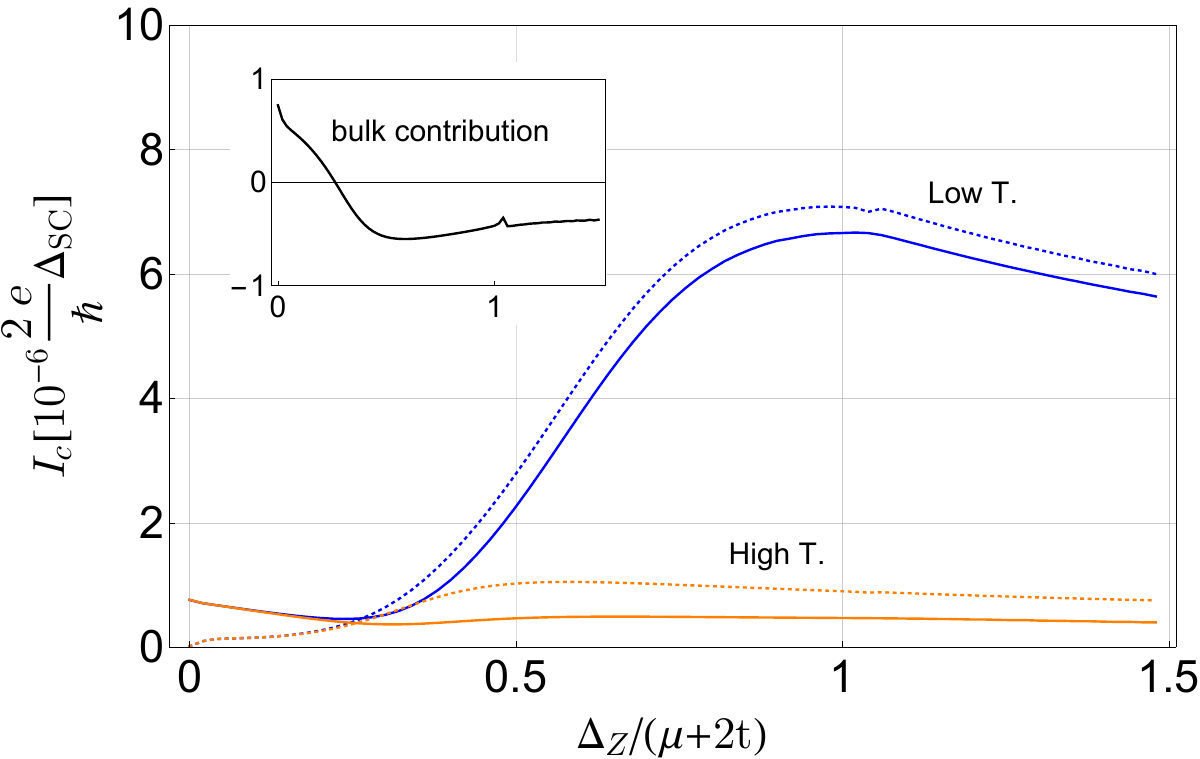}
\par\end{centering}
\caption{
(color online)
Zero bias critical current between a topological superconductor and the nanowire obtained by the tight-binding model is given.
Solid lines indicate critical current. Dotted lines are for current through the localized states.
\hlred{We take the parameters $\Delta_{sc} = 0.3, \Delta_{so} = 0.3, \Delta_{Z} = 1.5$ for the topological superconductors and $\Delta_{sc} = 0.3, \Delta_{so} = 0.3$ for the nanowire.
In both cases, the system comprises $N=500$ sites with $t=7/4$ and $\mu = -5/2$, thereby effectively setting the unit energy scale $\mu+2t=1$.
}
Tunneling amplitude is $\hlred{t_{\text{tunnel}}} = 0.02$.
Temperatures are $k_B T = 0.04, 0.005$..
A small kink near the phase transition, $\Delta_Z \approx 1.04$, is due to a systematic error, i.e. the localized state and bulk states are not well separated near the point.
(Inset) The critical current that bulk states contribute, i.e., bulk states to bulk states and localized states to bulk states.
}
\label{fig::cc:tsnw}
\end{figure*}

For the topological superconductor-nanowire junction (TS-NW), we \hlred{observe} that the majority of the supercurrent flows through the localized states (see Fig.~\ref{fig::cc:tsnw}).
Intriguingly, even when the nanowire is in the trivial phase, the subgap state cannot be disregarded; it makes a finite contribution to the critical current.
\hlred{In contrast to the NS-NW junction, the critical current is significantly affected by temperature since the predominant flow occurs through the lowest energy states, namely, the subgap state in the nanowire and the Majorana bound state in the topological superconductor.}

\section{Conclusion}

We have examined coupled one-dimensional topological subsystems\hlred{, which features the schematic coupling structure of a ladder form,} and each of which is in different topological states of matter. We reveal that this interaction results in a precursor effect during the topological phase transition within the entire system.
This effect is entirely governed by the symmetry classes of the subsystem Hamiltonians and the coupling term, and it is characterized by the continuous presence of subgap states within the bulk energy gap.

By analyzing the critical current in Josephson junctions incorporating topological superconductors, we have demonstrated the essential roles these subgap states play in the physical properties of nanostructures and low-dimensional materials.


\authorcontributions{
  Conceptualization, M.-S.C.;
  methodology, S.H. and M.-S.C.;
  validation, S.H. and M.-S.C.;
  formal analysis, S.H. and M.-S.C.;
  investigation, S.H. and M.-S.C.;
  data curation, M.-S.C.;
  writing---original draft preparation, S.H.;
  writing---review and editing, M.-S.C.;
  visualization, S.H. and M.-S.C.;
  supervision, M.-S.C.;
  project administration, M.-S.C.;
  funding acquisition, M.-S.C.
  All authors have read and agreed to the published version of the manuscript.}

\funding{This work was supported by the National Research Function (NRF) of Korea
(Grant Nos. 2022M3H3A106307411 and 2023R1A2C1005588) and by the Ministry of
Education through the BK21 Four program.}

\institutionalreview{Not applicable}

\informedconsent{Not applicable}

\dataavailability{All data created are included as figures in this paper.} 

\acknowledgments{Not applicable}

\conflictsofinterest{The authors declare no conflict of interest.}

\appendixtitles{yes} 
\appendixstart
\appendix

\section{Accidental symmetry of the two coupled Kitaev chains model}
\label{sec::appendix1}

In Section~\ref{sec::twokitaev}, we investigated the effects of inter-chain coupling with arbitrary phase.
Here, we will discuss the special case that the inter-chain pairing is pure imaginary.
The point is that this constraint on the inter-chain coupling can lead to additional anti-unitary symmetries which affect the topological class of the model.
However, such a constraint and result topological class is illusive in the sense that it easily breaks down in realistic situation such as local disorders and next nearest-neighbor (NNN) couplings.
On this ground, we emphasize that the classification and corresponding discussions in Section~\ref{sec::twokitaev} are physically valid in general.
Nevertheless, we present the analysis of the topological class of this special case to avoid potential confusion when one encounters the similar situation with illusive anti-unitary symmetries.

Recall that the phase of intra-chain pairings are fixed to $\Delta_1 = -|\Delta_1|$ and $\Delta_2 = |\Delta_2|$ without loss of generality; in other words, the phase of the inter-chain pairing is a gauge-fixed quantity.

With the given constraint (and assumption $t_{\text{int}}=0$), the model acquires a time-reversal-like symmetry operation $\hat{\mathcal{T}}$,
\begin{align}
\hat{\mathcal{T}} \hat{c}_{j,\alpha} \hat{\mathcal{T}}^{-1} &=
(-1)^\alpha \hat{c}_{j,\alpha}, \quad \alpha = 1,2
\nonumber \\
\hat{\mathcal{T}} i \hat{\mathcal{T}}^{-1} &= - i,
\end{align}
such that the Hamiltonian satisfies
\begin{align}
\hat{\mathcal{T}} \hat{H}_{\text{CKC}} \hat{\mathcal{T}}^{-1} &= \hat{H}_{\text{CKC}}.
\end{align}
It is stressed that this symmetry operation is \emph{not} related to the \emph{physical} time-reversal symmetry.
Since we consider the model of coupled Kitaev chains as an analogy to the nanowire model which is composed of electrons, the \emph{physical} time reversal symmetry operator should satisfy the relation $\hat{\mathcal{T}}_{\text{physical}}^2
 = (-1)^{\hat{N}}$, where $\hat{N}$ for electron number operator.
To the contrary, the symmetry operator $\hat{\mathcal{T}}$ of our interest here satisfies $\hat{\mathcal{T}}^2=1$.

\begin{figure}
\includegraphics[width=0.31\textwidth]{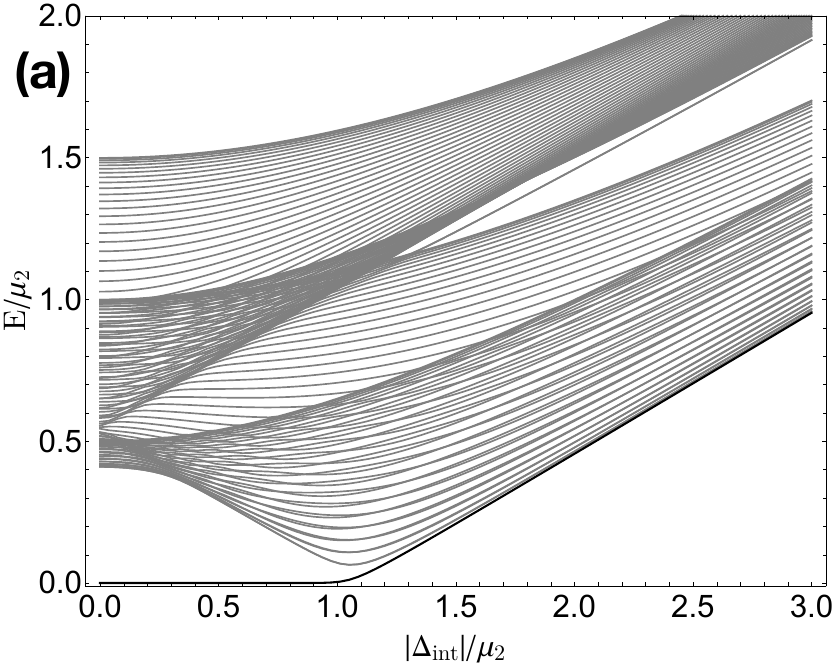} 
\includegraphics[width=0.31\textwidth]{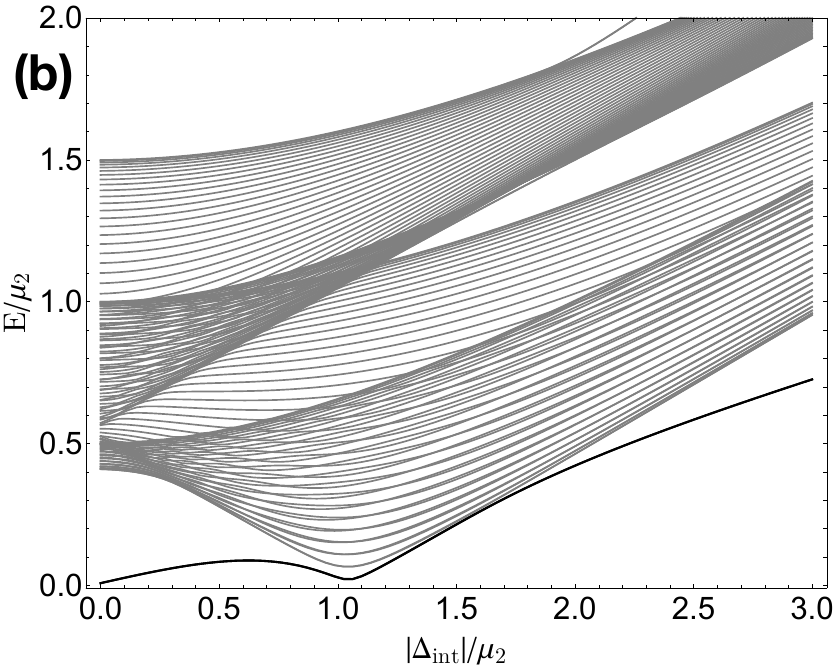} 
\includegraphics[width=0.31\textwidth]{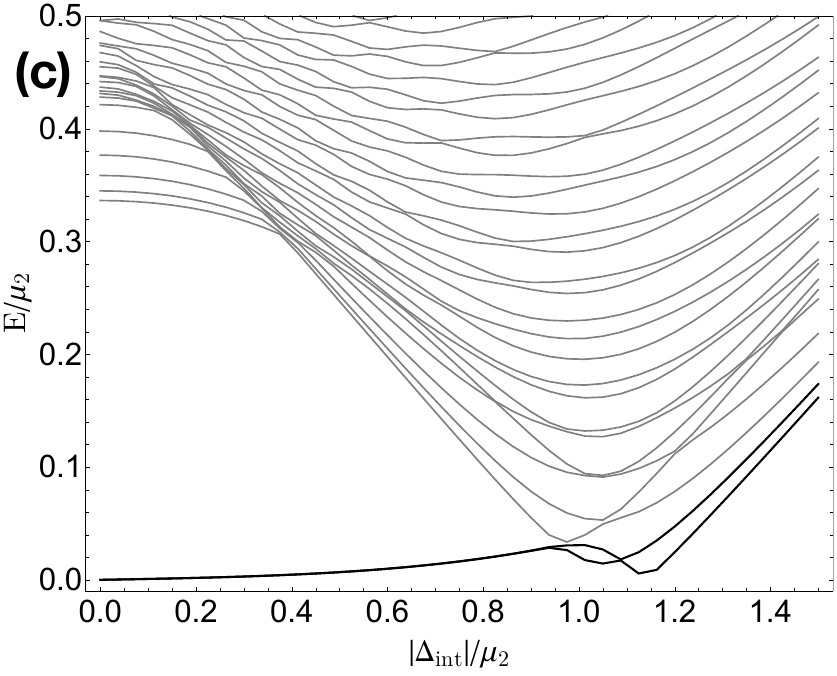} 
\caption{
Energy levels as a function $|\Delta_{\text{int}}|$, where $|\Delta_{\text{int}}|$ is pure imaginary.
(a) The ergy levels along the positive vertical axis of Fig.~\ref{fig::twokitaev:pairing} (a).
(b) The energy levels along the positive vertical axis Fig.~\ref{fig::twokitaev:total} (a).
(c) The energy levels when introducing a next-nearest-neighbor hopping term, $\sum_{j,\alpha} t^{(2)} \hat{c}_{j+2,\alpha}^{\dagger} \hat{c}_{j,\alpha} + h.c.$ where $t^{(2)} = 0.1 e^{i\pi/4} \mu_2 $.
Parameters: The number of sites for a chain $N=60$,
on-site energy $\mu_1=0$, $\mu_2=1$ (unit energy scale),
intra-chain pairings $\Delta_{1}=-\Delta_{2}=-\hlred{\mu_2/2}$,
hoppings \hlred{$t_1=t_2=\mu_2$.}
}
\label{fig::twokitaev:appendix}
\end{figure}

\hlred{While} the symmetry described by $\hat{\mathcal{T}}$ is not a \emph{physical} time-reversal symmetry, it determines the symmetry class of the Hamiltonian.
Together with the particle-hole constraint of the BdG Hamiltonian,
\hlred{the system falls into the BDI symmetry class, with the winding number as the relevant topological invariant.}
This winding number captures the number of the \emph{chiral Majorana bound states} localized at one end;
that is, it represents the number of Majorana bound states with different chirality (i.e., eigenvalues of the chiral operator) at one end.
In \hlred{the context of our model,} briefly mentioned in Sec.~\ref{sec::twokitaev},
if we change the strength of the inter-chain pairing $\Delta_{\text{int}}$, the two chiral Majoranas at each end remain at zero energy until the gap closes.
The winding number is $2$ before the gap closing, so the two Majorana bound states on the left end of the chains have chirality $-1$ while the other two states on the right end have chirality $+1$.
After the gap closing, the winding number is changed to 0 and there are no Majorana bound states anymore; see Fig.~\ref{fig::twokitaev:appendix} (a).

However, unlike the usual unpaired Majorana bound states, the energy of the above states can be easily lifted from zero as the time-reversal-like symmetry can be easily broken if we introduce additional interactions or local disorder.
For example, inter-chain hopping terms, $t_{\text{inter}} \hat{c}_{j,1}^{\dagger} \hat{c}_{j,2}$, or NNN hopping terms with nonzero complex phase, i.e. $t^{(2)} \hat{c}_{j+2,\alpha}^{\dagger}\hat{c}_{j,\alpha}$, would break the time-reversal-like symmetry, such that the energy of boundary states are lifted from zero and have finite values.

In Fig.~\ref{fig::twokitaev:appendix} (b), we consider the model with boundary hopping $t_{\text{int}} = |\Delta_{\text{int}}|$ turned on, which is already considered in Sec. \ref{sec::twocoupledkitaevchains} and Fig.~\ref{fig::twokitaev:total}.
As the inter-chain pairing $\Delta_{\text{int}}$ increases, the zero energy is lifted from zero before the band touching.
It is due to direct coupling between two chiral Majorana bound states via the boundary hopping term.
In Fig.~\ref{fig::twokitaev:appendix} (c), we introduce NNN hopping terms for each chain, $\sum_{j,\alpha} t^{(2)} \hat{c}_{j+2,\alpha}^{\dagger}\hat{c}_{j,\alpha}+ h.c.$ where $t^{(2)} = 0.1 e^{i\pi/4} \mu_2 $.
Though the NNN hopping term does not couple the two chiral Majorana bound states directly as it is intra-chain coupling, it changes the character of Majorana bound states itself for each chain, such that the inter-chain pairing couples the two Majorana bound states and gives them finite energy.
In the both cases, the additional terms break the time-reversal-like symmetry. The relevant symmetry classes of the systems are D and relevant topological invariants $Z_2$(Chern-Simon integral) are 0, i.e. the systems are in a topologically trivial phase.

By considering other constraints rather than pure imaginary $\Delta_{\text{int}}$, one may find similar time-reversal-like symmetry.
For example, without boundary hopping terms, the pure real inter-chain pairing implants a time-reversal like symmetry.
The detailed properties of the symmetry operation and corresponding winding number would change with the pure imaginary $\Delta_{\text{int}}$ case.
Another example is pure imaginary boundary hopping without any inter-chain pairings.

\begin{adjustwidth}{-\extralength}{0cm}

\reftitle{References}

\bibliography{main}

\begin{thebibliography}{999}

\bibitem[Goldenfeld(1992)]{Goldenfeld92a}
Goldenfeld, N.
\newblock {\em Phase Transitions and the Renormalization Group};
  Addison-Wesley: New York,  1992.

\bibitem[Hasan and Kane(2010)]{Hasan10a}
Hasan, M.Z.; Kane, C.L.
\newblock {Colloquium: Topological insulators}.
\newblock {\em Reviews of Modern Physics} {\bf 2010}, {\em 82},~3045--3067.
\newblock {\url{https://doi.org/10.1103/RevModPhys.82.3045}}.

\bibitem[Qi and Zhang(2011)]{Qi11a}
Qi, X.L.; Zhang, S.C.
\newblock Topological insulators and superconductors.
\newblock {\em Rev. Mod. Phys.} {\bf 2011}, {\em 83},~1057--1110.

\bibitem[Chiu et~al.(2016)Chiu, Teo, Schnyder, and Ryu]{Chiu16a}
Chiu, C.K.; Teo, J.C.Y.; Schnyder, A.P.; Ryu, S.
\newblock Classification of topological quantum matter with symmetries.
\newblock {\em Rev. Mod. Phys.} {\bf 2016}, {\em 88},~035005.
\newblock {\url{https://doi.org/10.1103/RevModPhys.88.035005}}.

\bibitem[Armitage et~al.(2018)Armitage, Mele, and Vishwanath]{Armitage18a}
Armitage, N.; Mele, E.; Vishwanath, A.
\newblock Weyl and Dirac semimetals in three-dimensional solids.
\newblock {\em Reviews of Modern Physics} {\bf 2018}, {\em 90},~015001.

\bibitem[Aguado(2017)]{Aguado17}
Aguado, R.
\newblock Majorana quasiparticles in condensed matter.
\newblock {\em La Rivista del Nuovo Cimento} {\bf 2017}, {\em 40},~523--593.
\newblock {\url{https://doi.org/10.1393/ncr/i2017-10141-9}}.

\bibitem[Haim et~al.(2015{\natexlab{a}})Haim, Berg, von Oppen, and
  Oreg]{haim15}
Haim, A.; Berg, E.; von Oppen, F.; Oreg, Y.
\newblock Current correlations in a Majorana beam splitter.
\newblock {\em Phys. Rev. B} {\bf 2015}, {\em 92},~245112.
\newblock {\url{https://doi.org/10.1103/PhysRevB.92.245112}}.

\bibitem[Haim et~al.(2015{\natexlab{b}})Haim, Berg, von Oppen, and
  Oreg]{haim15b}
Haim, A.; Berg, E.; von Oppen, F.; Oreg, Y.
\newblock Signatures of Majorana Zero Modes in Spin-Resolved Current
  Correlations.
\newblock {\em Phys. Rev. Lett.} {\bf 2015}, {\em 114},~166406.
\newblock {\url{https://doi.org/10.1103/PhysRevLett.114.166406}}.

\bibitem[Tiira et~al.(2017)Tiira, Strambini, Amado, Roddaro, San-Jose, Aguado,
  Bergeret, Ercolani, Sorba, and Giazotto]{Tiira17a}
Tiira, J.; Strambini, E.; Amado, M.; Roddaro, S.; San-Jose, P.; Aguado, R.;
  Bergeret, F.S.; Ercolani, D.; Sorba, L.; Giazotto, F.
\newblock {Magnetically-driven colossal supercurrent enhancement in InAs
  nanowire Josephson junctions}.
\newblock {\em Nature Communications} {\bf 2017}, {\em 8},~14984.

\bibitem[G{\"u}l et~al.(2018)G{\"u}l, Zhang, Bommer, de~Moor, Car, Plissard,
  Bakkers, Geresdi, Watanabe, Taniguchi, and Kouwenhoven]{Onder18a}
G{\"u}l, {\"O}.; Zhang, H.; Bommer, J.D.S.; de~Moor, M.W.A.; Car, D.; Plissard,
  S.R.; Bakkers, E.P.A.M.; Geresdi, A.; Watanabe, K.; Taniguchi, T.;  et~al.
\newblock Ballistic Majorana nanowire devices.
\newblock {\em Nature Nanotechnology} {\bf 2018}, {\em 13},~192--197.
\newblock {\url{https://doi.org/10.1038/s41565-017-0032-8}}.

\bibitem[Ren et~al.(2021)Ren, Ke, Guo, Zhang, and L\"u]{Ren21a}
Ren, J.T.; Ke, S.S.; Guo, Y.; Zhang, H.W.; L\"u, H.F.
\newblock Phase diagram and quantum transport in a semiconductor-superconductor
  hybrid nanowire with long-range pairing interactions.
\newblock {\em Phys. Rev. B} {\bf 2021}, {\em 103},~045428.
\newblock {\url{https://doi.org/10.1103/PhysRevB.103.045428}}.

\bibitem[Levajac et~al.(2023)Levajac, Wang, Sfiligoj, Lemang, Wolff, Bordin,
  Badawy, Gazibegovic, Bakkers, and Kouwenhoven]{Levajac2023}
Levajac, V.; Wang, J.Y.; Sfiligoj, C.; Lemang, M.; Wolff, J.C.; Bordin, A.;
  Badawy, G.; Gazibegovic, S.; Bakkers, E.P.A.M.; Kouwenhoven, L.P.
\newblock Subgap spectroscopy along hybrid nanowires by nm-thick tunnel
  barriers.
\newblock {\em Nature Communications} {\bf 2023}, {\em 14},~6647.
\newblock {\url{https://doi.org/10.1038/s41467-023-42422-z}}.

\bibitem[Liu et~al.(2012)Liu, Potter, Law, and Lee]{liu12}
Liu, J.; Potter, A.C.; Law, K.T.; Lee, P.A.
\newblock Zero-Bias Peaks in the Tunneling Conductance of Spin-Orbit-Coupled
  Superconducting Wires with and without Majorana End-States.
\newblock {\em Phys. Rev. Lett.} {\bf 2012}, {\em 109},~267002.
\newblock {\url{https://doi.org/10.1103/PhysRevLett.109.267002}}.

\bibitem[Huang et~al.(2018)Huang, Sau, Stanescu, and Das~Sarma]{huang18}
Huang, Y.; Sau, J.D.; Stanescu, T.D.; Das~Sarma, S.
\newblock Quasiparticle gaps in multiprobe Majorana nanowires.
\newblock {\em Phys. Rev. B} {\bf 2018}, {\em 98},~224512.
\newblock {\url{https://doi.org/10.1103/PhysRevB.98.224512}}.

\bibitem[Bagrets and Altland(2012)]{Bagrets12a}
Bagrets, D.; Altland, A.
\newblock Class $D$ Spectral Peak in Majorana Quantum Wires.
\newblock {\em Phys. Rev. Lett.} {\bf 2012}, {\em 109},~227005.
\newblock {\url{https://doi.org/10.1103/PhysRevLett.109.227005}}.

\bibitem[Pe\~naranda et~al.(2018)Pe\~naranda, Aguado, San-Jose, and
  Prada]{Penaranda18a}
Pe\~naranda, F.; Aguado, R.; San-Jose, P.; Prada, E.
\newblock Quantifying wave-function overlaps in inhomogeneous Majorana
  nanowires.
\newblock {\em Phys. Rev. B} {\bf 2018}, {\em 98},~235406.
\newblock {\url{https://doi.org/10.1103/PhysRevB.98.235406}}.

\bibitem[Huang et~al.(2018)Huang, Pan, Liu, Sau, Stanescu, and
  Das~Sarma]{huang18c}
Huang, Y.; Pan, H.; Liu, C.X.; Sau, J.D.; Stanescu, T.D.; Das~Sarma, S.
\newblock Metamorphosis of Andreev bound states into Majorana bound states in
  pristine nanowires.
\newblock {\em Phys. Rev. B} {\bf 2018}, {\em 98},~144511.
\newblock {\url{https://doi.org/10.1103/PhysRevB.98.144511}}.

\bibitem[Prada et~al.(2012)Prada, San-Jose, and Aguado]{prada12}
Prada, E.; San-Jose, P.; Aguado, R.
\newblock Transport spectroscopy of $NS$ nanowire junctions with Majorana
  fermions.
\newblock {\em Phys. Rev. B} {\bf 2012}, {\em 86},~180503.
\newblock {\url{https://doi.org/10.1103/PhysRevB.86.180503}}.

\bibitem[Kells et~al.(2012)Kells, Meidan, and Brouwer]{Kells12a}
Kells, G.; Meidan, D.; Brouwer, P.W.
\newblock Near-zero-energy end states in topologically trivial spin-orbit
  coupled superconducting nanowires with a smooth confinement.
\newblock {\em Phys. Rev. B} {\bf 2012}, {\em 86},~100503.
\newblock {\url{https://doi.org/10.1103/PhysRevB.86.100503}}.

\bibitem[DeGottardi et~al.(2013)DeGottardi, Sen, and
  Vishveshwara]{degottardi13}
DeGottardi, W.; Sen, D.; Vishveshwara, S.
\newblock Majorana Fermions in Superconducting 1D Systems Having Periodic,
  Quasiperiodic, and Disordered Potentials.
\newblock {\em Phys. Rev. Lett.} {\bf 2013}, {\em 110},~146404.
\newblock {\url{https://doi.org/10.1103/PhysRevLett.110.146404}}.

\bibitem[Ojanen(2013)]{ojanen13}
Ojanen, T.
\newblock Topological $\ensuremath{\pi}$ Josephson junction in superconducting
  Rashba wires.
\newblock {\em Phys. Rev. B} {\bf 2013}, {\em 87},~100506.
\newblock {\url{https://doi.org/10.1103/PhysRevB.87.100506}}.

\bibitem[Roy et~al.(2013)Roy, Bondyopadhaya, and Tewari]{roy13}
Roy, D.; Bondyopadhaya, N.; Tewari, S.
\newblock Topologically trivial zero-bias conductance peak in semiconductor
  Majorana wires from boundary effects.
\newblock {\em Phys. Rev. B} {\bf 2013}, {\em 88},~020502.
\newblock {\url{https://doi.org/10.1103/PhysRevB.88.020502}}.

\bibitem[Rainis et~al.(2013)Rainis, Trifunovic, Klinovaja, and Loss]{rainis13}
Rainis, D.; Trifunovic, L.; Klinovaja, J.; Loss, D.
\newblock Towards a realistic transport modeling in a superconducting nanowire
  with Majorana fermions.
\newblock {\em Phys. Rev. B} {\bf 2013}, {\em 87},~024515.
\newblock {\url{https://doi.org/10.1103/PhysRevB.87.024515}}.

\bibitem[Adagideli et~al.(2014)Adagideli, Wimmer, and Teker]{adagideli14}
Adagideli, i.d.I.; Wimmer, M.; Teker, A.
\newblock Effects of electron scattering on the topological properties of
  nanowires: Majorana fermions from disorder and superlattices.
\newblock {\em Phys. Rev. B} {\bf 2014}, {\em 89},~144506.
\newblock {\url{https://doi.org/10.1103/PhysRevB.89.144506}}.

\bibitem[Stanescu et~al.(2014)Stanescu, Lutchyn, and Das~Sarma]{stanescu14}
Stanescu, T.D.; Lutchyn, R.M.; Das~Sarma, S.
\newblock Soft superconducting gap in semiconductor-based Majorana nanowires.
\newblock {\em Phys. Rev. B} {\bf 2014}, {\em 90},~085302.
\newblock {\url{https://doi.org/10.1103/PhysRevB.90.085302}}.

\bibitem[Klinovaja and Loss(2015)]{jelena15}
Klinovaja, J.; Loss, D.
\newblock Fermionic and Majorana bound states in hybrid nanowires with
  non-uniform spin-orbit interaction.
\newblock {\em The European Physical Journal B} {\bf 2015}, {\em 88},~62.
\newblock {\url{https://doi.org/10.1140/epjb/e2015-50882-2}}.

\bibitem[San-Jose et~al.(2016)San-Jose, Cayao, Prada, and Aguado]{sanjose16}
San-Jose, P.; Cayao, J.; Prada, E.; Aguado, R.
\newblock Majorana bound states from exceptional points in non-topological
  superconductors.
\newblock {\em Scientific Reports} {\bf 2016}, {\em 6},~21427.
\newblock {\url{https://doi.org/10.1038/srep21427}}.

\bibitem[Deng et~al.(2016)Deng, Vaitiek{\.e}nas, Hansen, Danon, Leijnse,
  Flensberg, Nyg{\aa}rd, Krogstrup, and Marcus]{mtdeng16}
Deng, M.T.; Vaitiek{\.e}nas, S.; Hansen, E.B.; Danon, J.; Leijnse, M.;
  Flensberg, K.; Nyg{\aa}rd, J.; Krogstrup, P.; Marcus, C.M.
\newblock Majorana bound state in a coupled quantum-dot hybrid-nanowire system.
\newblock {\em Science} {\bf 2016}, {\em 354},~1557--1562,
  \href{http://xxx.lanl.gov/abs/https://www.science.org/doi/pdf/10.1126/science.aaf3961}{{\normalfont
  [https://www.science.org/doi/pdf/10.1126/science.aaf3961]}}.
\newblock {\url{https://doi.org/10.1126/science.aaf3961}}.

\bibitem[Liu et~al.(2017)Liu, Sau, Stanescu, and Das~Sarma]{liu17}
Liu, C.X.; Sau, J.D.; Stanescu, T.D.; Das~Sarma, S.
\newblock Andreev bound states versus Majorana bound states in quantum
  dot-nanowire-superconductor hybrid structures: Trivial versus topological
  zero-bias conductance peaks.
\newblock {\em Phys. Rev. B} {\bf 2017}, {\em 96},~075161.
\newblock {\url{https://doi.org/10.1103/PhysRevB.96.075161}}.

\bibitem[Ptok et~al.(2017)Ptok, Kobia\l{}ka, and Doma\ifmmode~\acute{n}\else
  \'{n}\fi{}ski]{ptok17}
Ptok, A.; Kobia\l{}ka, A.; Doma\ifmmode~\acute{n}\else \'{n}\fi{}ski, T.
\newblock Controlling the bound states in a quantum-dot hybrid nanowire.
\newblock {\em Phys. Rev. B} {\bf 2017}, {\em 96},~195430.
\newblock {\url{https://doi.org/10.1103/PhysRevB.96.195430}}.

\bibitem[Fleckenstein et~al.(2018)Fleckenstein, Dom\'{\i}nguez, Traverso~Ziani,
  and Trauzettel]{fleckenstein18}
Fleckenstein, C.; Dom\'{\i}nguez, F.; Traverso~Ziani, N.; Trauzettel, B.
\newblock Decaying spectral oscillations in a Majorana wire with finite
  coherence length.
\newblock {\em Phys. Rev. B} {\bf 2018}, {\em 97},~155425.
\newblock {\url{https://doi.org/10.1103/PhysRevB.97.155425}}.

\bibitem[Avila et~al.(2019)Avila, Pe{\~n}aranda, Prada, San-Jose, and
  Aguado]{avila19}
Avila, J.; Pe{\~n}aranda, F.; Prada, E.; San-Jose, P.; Aguado, R.
\newblock Non-hermitian topology as a unifying framework for the Andreev versus
  Majorana states controversy.
\newblock {\em Communications Physics} {\bf 2019}, {\em 2},~133.
\newblock {\url{https://doi.org/10.1038/s42005-019-0231-8}}.

\bibitem[Moore et~al.(2018{\natexlab{a}})Moore, Zeng, Stanescu, and
  Tewari]{moore18}
Moore, C.; Zeng, C.; Stanescu, T.D.; Tewari, S.
\newblock Quantized zero-bias conductance plateau in
  semiconductor-superconductor heterostructures without topological Majorana
  zero modes.
\newblock {\em Phys. Rev. B} {\bf 2018}, {\em 98},~155314.
\newblock {\url{https://doi.org/10.1103/PhysRevB.98.155314}}.

\bibitem[Moore et~al.(2018{\natexlab{b}})Moore, Stanescu, and Tewari]{moore18c}
Moore, C.; Stanescu, T.D.; Tewari, S.
\newblock Two-terminal charge tunneling: Disentangling Majorana zero modes from
  partially separated Andreev bound states in semiconductor-superconductor
  heterostructures.
\newblock {\em Phys. Rev. B} {\bf 2018}, {\em 97},~165302.
\newblock {\url{https://doi.org/10.1103/PhysRevB.97.165302}}.

\bibitem[Stanescu and Tewari(2019)]{tudor19}
Stanescu, T.D.; Tewari, S.
\newblock Robust low-energy Andreev bound states in
  semiconductor-superconductor structures: Importance of partial separation of
  component Majorana bound states.
\newblock {\em Phys. Rev. B} {\bf 2019}, {\em 100},~155429.
\newblock {\url{https://doi.org/10.1103/PhysRevB.100.155429}}.

\bibitem[Woods et~al.(2019)Woods, Chen, Frolov, and Stanescu]{woods19}
Woods, B.D.; Chen, J.; Frolov, S.M.; Stanescu, T.D.
\newblock Zero-energy pinning of topologically trivial bound states in
  multiband semiconductor-superconductor nanowires.
\newblock {\em Phys. Rev. B} {\bf 2019}, {\em 100},~125407.
\newblock {\url{https://doi.org/10.1103/PhysRevB.100.125407}}.

\bibitem[Chen et~al.(2019)Chen, Woods, Yu, Hocevar, Car, Plissard, Bakkers,
  Stanescu, and Frolov]{chen19}
Chen, J.; Woods, B.D.; Yu, P.; Hocevar, M.; Car, D.; Plissard, S.R.; Bakkers,
  E.P.A.M.; Stanescu, T.D.; Frolov, S.M.
\newblock Ubiquitous Non-Majorana Zero-Bias Conductance Peaks in Nanowire
  Devices.
\newblock {\em Phys. Rev. Lett.} {\bf 2019}, {\em 123},~107703.
\newblock {\url{https://doi.org/10.1103/PhysRevLett.123.107703}}.

\bibitem[Awoga et~al.(2019)Awoga, Cayao, and Black-Schaffer]{awoga19}
Awoga, O.A.; Cayao, J.; Black-Schaffer, A.M.
\newblock Supercurrent Detection of Topologically Trivial Zero-Energy States in
  Nanowire Junctions.
\newblock {\em Phys. Rev. Lett.} {\bf 2019}, {\em 123},~117001.
\newblock {\url{https://doi.org/10.1103/PhysRevLett.123.117001}}.

\bibitem[Prada et~al.(2020)Prada, San-Jose, de~Moor, Geresdi, Lee, Klinovaja,
  Loss, Nyg{\aa}rd, Aguado, and Kouwenhoven]{Prada2020}
Prada, E.; San-Jose, P.; de~Moor, M.W.A.; Geresdi, A.; Lee, E.J.H.; Klinovaja,
  J.; Loss, D.; Nyg{\aa}rd, J.; Aguado, R.; Kouwenhoven, L.P.
\newblock From Andreev to Majorana bound states in hybrid
  superconductor--semiconductor nanowires.
\newblock {\em Nature Reviews Physics} {\bf 2020}, {\em 2},~575--594.
\newblock {\url{https://doi.org/10.1038/s42254-020-0228-y}}.

\bibitem[Ahn et~al.(2021)Ahn, Pan, Woods, Stanescu, and Das~Sarma]{ahn21}
Ahn, S.; Pan, H.; Woods, B.; Stanescu, T.D.; Das~Sarma, S.
\newblock Estimating disorder and its adverse effects in semiconductor Majorana
  nanowires.
\newblock {\em Phys. Rev. Mater.} {\bf 2021}, {\em 5},~124602.
\newblock {\url{https://doi.org/10.1103/PhysRevMaterials.5.124602}}.

\bibitem[Marra and Nigro(2022)]{marra22}
Marra, P.; Nigro, A.
\newblock Majorana/Andreev crossover and the fate of the topological phase
  transition in inhomogeneous nanowires.
\newblock {\em Journal of Physics: Condensed Matter} {\bf 2022}, {\em
  34},~124001.
\newblock {\url{https://doi.org/10.1088/1361-648X/ac44d2}}.

\bibitem[Fukaya et~al.(2023)Fukaya, Yada, Tanaka, Gentile, and Cuoco]{fukaya23}
Fukaya, Y.; Yada, K.; Tanaka, Y.; Gentile, P.; Cuoco, M.
\newblock Particle-hole spectral asymmetry at the edge of multiorbital
  noncentrosymmetric superconductors.
\newblock {\em Phys. Rev. B} {\bf 2023}, {\em 108},~L020502.
\newblock {\url{https://doi.org/10.1103/PhysRevB.108.L020502}}.

\bibitem[van Driel et~al.(2023)van Driel, Wang, Bordin, van Loo, Zatelli,
  Mazur, Xu, Gazibegovic, Badawy, Bakkers, Kouwenhoven, and Dvir]{Driel23}
van Driel, D.; Wang, G.; Bordin, A.; van Loo, N.; Zatelli, F.; Mazur, G.P.; Xu,
  D.; Gazibegovic, S.; Badawy, G.; Bakkers, E.P.A.M.;  et~al.
\newblock Spin-filtered measurements of Andreev bound states in
  semiconductor-superconductor nanowire devices.
\newblock {\em Nature Communications} {\bf 2023}, {\em 14},~6880.
\newblock {\url{https://doi.org/10.1038/s41467-023-42026-7}}.

\bibitem[Hess et~al.(2023)Hess, Legg, Loss, and Klinovaja]{hessDloss23}
Hess, R.; Legg, H.F.; Loss, D.; Klinovaja, J.
\newblock Trivial Andreev Band Mimicking Topological Bulk Gap Reopening in the
  Nonlocal Conductance of Long Rashba Nanowires.
\newblock {\em Phys. Rev. Lett.} {\bf 2023}, {\em 130},~207001.
\newblock {\url{https://doi.org/10.1103/PhysRevLett.130.207001}}.

\bibitem[Wang et~al.(2018)Wang, Shao, Zhao, Sheng, Wang, and
  Xing]{Huaiqiang18a}
Wang, H.; Shao, L.B.; Zhao, Y.X.; Sheng, L.; Wang, B.G.; Xing, D.Y.
\newblock Effective long-range pairing and hopping in topological nanowires
  weakly coupled to $s$-wave superconductors.
\newblock {\em Phys. Rev. B} {\bf 2018}, {\em 98},~174512.
\newblock {\url{https://doi.org/10.1103/PhysRevB.98.174512}}.

\bibitem[Reeg et~al.(2017)Reeg, Loss, and Klinovaja]{reegDloss2017}
Reeg, C.; Loss, D.; Klinovaja, J.
\newblock Finite-size effects in a nanowire strongly coupled to a thin
  superconducting shell.
\newblock {\em Phys. Rev. B} {\bf 2017}, {\em 96},~125426.
\newblock {\url{https://doi.org/10.1103/PhysRevB.96.125426}}.

\bibitem[Anselmetti et~al.(2019)Anselmetti, Martinez, M\'enard, Puglia,
  Malinowski, Lee, Choi, Pendharkar, Palmstr\o{}m, Marcus, Casparis, and
  Higginbotham]{anselmetti19}
Anselmetti, G.L.R.; Martinez, E.A.; M\'enard, G.C.; Puglia, D.; Malinowski,
  F.K.; Lee, J.S.; Choi, S.; Pendharkar, M.; Palmstr\o{}m, C.J.; Marcus, C.M.;
  et~al.
\newblock End-to-end correlated subgap states in hybrid nanowires.
\newblock {\em Phys. Rev. B} {\bf 2019}, {\em 100},~205412.
\newblock {\url{https://doi.org/10.1103/PhysRevB.100.205412}}.

\bibitem[Alicea et~al.(2011)Alicea, Oreg, Refael, von Oppen, and
  Fisher]{Alicea11a}
Alicea, J.; Oreg, Y.; Refael, G.; von Oppen, F.; Fisher, M.P.A.
\newblock Non-Abelian statistics and topological quantum information processing
  in 1D wire networks.
\newblock {\em Nat Phys} {\bf 2011}, {\em 7},~412--417.

\bibitem[Kitaev(2006)]{Kitaev06b}
Kitaev, A.
\newblock Anyons in an exactly solved model and beyond.
\newblock {\em Ann. Phys.} {\bf 2006}, {\em 321},~2 -- 111.

\bibitem[Zhang et~al.(2021)Zhang, Liu, Gazibegovic, Xu, Logan, Wang, van Loo,
  Bommer, de~Moor, Car, het Veld, van Veldhoven, Koelling, Verheijen,
  Pendharkar, Pennachio, Shojaei, Lee, Palmstr{\o}m, Bakkers, Sarma, and
  Kouwenhoven]{Zhang21a}
Zhang, H.; Liu, C.X.; Gazibegovic, S.; Xu, D.; Logan, J.A.; Wang, G.; van Loo,
  N.; Bommer, J.D.S.; de~Moor, M.W.A.; Car, D.;  et~al.
\newblock Retraction Note: Quantized Majorana conductance.
\newblock {\em Nature} {\bf 2021}, {\em 591},~E30--E30.
\newblock {\url{https://doi.org/10.1038/s41586-021-03373-x}}.

\bibitem[Kitaev(2001)]{Kitaev01a}
Kitaev, A.Y.
\newblock Unpaired Majorana fermions in quantum wires.
\newblock {\em Physics-Uspekhi} {\bf 2001}, {\em 44},~131.
\newblock {\url{https://doi.org/10.1070/1063-7869/44/10S/S29}}.

\bibitem[Oreg et~al.(2010)Oreg, Refael, and von Oppen]{Oreg10a}
Oreg, Y.; Refael, G.; von Oppen, F.
\newblock Helical Liquids and Majorana Bound States in Quantum Wires.
\newblock {\em Phys. Rev. Lett.} {\bf 2010}, {\em 105},~177002.
\newblock {\url{https://doi.org/10.1103/PhysRevLett.105.177002}}.

\bibitem[Lutchyn et~al.(2010)Lutchyn, Sau, and Das~Sarma]{Lutchyn10a}
Lutchyn, R.M.; Sau, J.D.; Das~Sarma, S.
\newblock Majorana Fermions and a Topological Phase Transition in
  Semiconductor-Superconductor Heterostructures.
\newblock {\em Phys. Rev. Lett.} {\bf 2010}, {\em 105},~077001.
\newblock {\url{https://doi.org/10.1103/PhysRevLett.105.077001}}.

\bibitem[Zazunov and Egger(2012)]{Zazunov12a}
Zazunov, A.; Egger, R.
\newblock Supercurrent blockade in Josephson junctions with a Majorana wire.
\newblock {\em Phys. Rev. B} {\bf 2012}, {\em 85},~104514.
\newblock {\url{https://doi.org/10.1103/PhysRevB.85.104514}}.

\end{thebibliography}

\PublishersNote{}
\end{adjustwidth}
\end{document}